\documentclass[conference]{IEEEtran}
\IEEEoverridecommandlockouts
\usepackage{cite}
\usepackage{amsmath,amssymb,amsfonts}
\usepackage{algorithmic}
\usepackage{graphicx}
\usepackage{textcomp}
\usepackage{xcolor}
\usepackage{hyperref}
\usepackage{tabularx}
\usepackage{tabulary}
\usepackage{ltablex}
\usepackage{ragged2e}
\usepackage{caption}

\def\BibTeX{{\rm B\kern-.05em{\sc i\kern-.025em b}\kern-.08em
    T\kern-.1667em\lower.7ex\hbox{E}\kern-.125emX}}
\usepackage{supertabular}
\begin{document}

\title{Evaluating the Performance of Machine Learning Algorithms in Financial Market Forecasting: A Comprehensive Survey 
\\}

\author{\IEEEauthorblockN{Lukas Ryll*}
\IEEEauthorblockA{\textit{Centre for Alternative Finance} \\
\textit{University of Cambridge}\\
Cambridge, UK \\
lr487@cam.ac.uk}\thanks{*Corresponding author}
\and
\IEEEauthorblockN{Sebastian Seidens}
\IEEEauthorblockA{\textit{Allianz Endowed Chair of Finance} \\
\textit{WHU - Otto Beisheim School of Management}\\
Vallendar, Germany \\
sebastian.seidens@whu.edu}
}

\maketitle

\begin{abstract}
With increasing competition and pace in the financial markets, robust forecasting methods are becoming more and more valuable to investors. While machine learning algorithms offer a proven way of modeling non-linearities in time series, their advantages against common stochastic models in the domain of financial market prediction are largely based on limited empirical results. The same holds true for determining advantages of certain machine learning architectures against others. \\This study surveys more than 150 related articles on applying machine learning to financial market forecasting. Based on a comprehensive literature review, we build a table across seven main parameters describing the experiments conducted in these studies. Through listing and classifying different algorithms, we also introduce a simple, standardized syntax for textually representing machine learning algorithms. Based on performance metrics gathered from papers included in the survey, we further conduct rank analyses to assess the comparative performance of different algorithm classes. 
\\Our analysis shows that machine learning algorithms tend to outperform most traditional stochastic methods in financial market forecasting. We further find evidence that, on average, recurrent neural networks outperform feed forward neural networks as well as support vector machines which implies the existence of exploitable temporal dependencies in financial time series across multiple asset classes and geographies.
\end{abstract}

\vspace{2mm}
\section{Introduction}
Since the early beginnings of capital markets, investors have tried to gain a competitive advantage over other market participants, and being able to accurately predict time series undoubtedly represents a constant topic of interest for market participants. Given the growth in available data sources and the increasing interconnectedness of investors, fast and efficient decision making is becoming more important than ever. Machine learning algorithms offer capabilities in approximating non-linear functions, dealing with noisy, non-stationary data, and discovering latent patterns in datasets. \\
With advances in machine learning throughout the last decades, most notably tackling issues arising from gradient flow which made recurrent networks impractical \cite{LSTM, GRU}, as well as significant progress in efficient computing using tensor operations on GPUs, machine learning algorithms pose a highly attractive option for financial time series forecasting.
Yet, despite the fast-growing importance of machine learning in the financial industry, the degree of academic consolidation and standardization in this field is still comparably sparse. Notwithstanding an increasing number of papers being released within this area of research over the course of the late 20th- and early 21st century, the literature currently fails to provide a compelling analysis of the different algorithms and their respective findings. \\Therefore, our study conducts a comprehensive, systematic review of existing works on trading algorithms to close this gap in contemporary research. Apart from providing an overview over the evolution of research in the application of machine learning in financial markets, this paper also suggests and confirms robust hypotheses about the performance of certain classes of algorithms based on rank analyses. For a comparison between different machine learning models through direct application, one would have to compile vast amounts of data from different exchanges and implement a large variety of different trading strategies. By gathering a large number of samples from different experimental methodologies, our study avoids capturing biases from authors using different financial interfaces and datasets and, thus, converges towards representing true differences between the actual algorithm classes.

In regard to the rank analyses, our main research hypothesis states that machine learning algorithms offer superior predictive performance to stochastic models due to their ability to capture recurring non-linear patterns in time series. As most modern supervised machine learning algorithms are trained using cross-validation, the resulting forecasts remain ‘smooth’, i.e., generalizable enough to avoid overfitting on the training data set, while still taking into account non-linearities. 
We further expect recurrent machine learning algorithms to systematically outperform purely feed-forward models in time series forecasting given their potential to model temporal dynamics, i.e., long-term dependencies within time series.\smallskip

The remainder of this paper is organized as follows. Section II provides a brief introduction to machine learning in financial market prediction while section III reviews existing literature on surveys and meta-studies in this field. Section IV outlines the research methodology and provides summary statistics of the dataset. Section V presents the findings of our analysis across measures, markets and time. Lastly, section VI concludes and presents some challenges for future research, followed by the table of studies.
\vspace{2mm}
\section{Machine learning in financial time series prediction}
While the term \textit{Machine Learning} remains ill-defined to some degree in contemporary literature, it can be broadly referred to as a process where a system interacts with its environment in such way that the structure of the system changes, and that this interaction process itself changes as a consequence to structural alterations. This is an abridged modification of a definition coined by \cite{Mendel} which was applied to the concept of neural networks by \cite{Haykin}. Within this high-level theorem, there are three main learning paradigms which each having different application areas in financial time series prediction.

Supervised learning is used for prediction tasks where a dataset with inputs and labeled targets is available. This may, for instance, entail using technical market indicators to predict whether the next day's stock price will go up (1) or down (0) (binary classification). Apart from classification, supervised learning algorithms may also perform regression tasks, i.e. predicting a continuous value instead of a class label. Taking the stock price example from above, this would translate to predicting the actual stock price or return instead of labeling winners and losers.
\\
Based on the results from forecasting or classification, there are several choices of financial interface, including building portfolios in a multi-asset classification/forecasting task \cite{Steiner}, systematic timing strategies \cite{Tenti} or simpler buy-and-hold strategies for single asset experiments (which can be found in the majority of all studies which we include in our survey).
\\
Unsupervised learning algorithms are usually designed for tasks that precede supervised learning, for instance, clustering or dimensionality reduction. An unsupervised learning algorithm may, for example, cluster stocks according to the similarity of their input features. The resulting cluster can then be further used for supervised classification \cite{Huang1}.
\\
Reinforcement learning is radically different from the two aforementioned paradigms in that it is based on an action-response model. Reinforcement learning algorithms learn certain action policies which maximize expected rewards. Thus, they are highly applicable to environments where actions and rewards are clearly defined, such as board games. The reinforcement learning process is commonly based on a value function which expresses the expected reward for an action undertaken at the current state of the system. In stock market forecasting, finding a suitable value function represents a major challenge, which is why other approaches, such as direct reinforcement using differential metric optimization objectives, have been proposed \cite{direct_rl}.
\\
The application of machine learning algorithms to financial time series has been covered by a large range of authors throughout the last two decades. Stemming from the simplest multi-layer perceptrons, state-of-the-art deep learning algorithms have evolved to capture time dynamics through recurrent neural architectures, and, specifically, gated neuron designs which allow for capturing long-term dependencies in time series (e.g., Long Short-Term Memory (LSTM) \cite{LSTM}). 
\\
Yet, while machine learning techniques are well suited for a variety of approximation tasks, they represent so-called 'black-box' models, meaning that their output behavior cannot be fully explained. In an on-line learning context, this property implies a lack of decision transparency which is essential for interpreting individual model outputs. This characteristic is especially vital in the case of abnormal market movements as the forecasting error may increase sharply for outlier events. Therefore, standardization and transparency in financial machine learning research are pivotal in illustrating varying behaviors across asset- and algorithm classes. 
\vspace{2mm}
\section{Literature}

As aforementioned, while existing research covers a variety of different algorithms, inputs, and concepts, there are few examples of studies which attempt to systematically review and compare existing works.
\cite{atsalakis} present a list of soft computing methods (including machine learning, evolutionary computing, and fuzzy logic) used in various research papers on trading algorithms. Their study largely serves as a passive reference due to its limited scope of analysis. While they conclude that soft computing algorithms represent a feasible stock forecasting method, they also note that ``[...] difficulties arise when defining the structure of the model (the hidden layers the neurons etc.). For the time being, the structure of the model is a matter of trial and error procedures.".
\\
A highly comprehensive perspective is provided by \cite{cavalcante} who present a brief overview of applications of computational intelligence to financial data in studies from 2009-2015. Apart from the survey, the paper establishes a standardized framework for constructing these algorithms. \cite{bahrammirzaee} presents similar results, concluding that artificial intelligence algorithms generally possess a higher accuracy than comparable statistical methods. Nevertheless, his study denies evidence of outperformance on an absolute scale. Our study addresses this doubt with a ranking analysis and finds significant evidence of the outperformance of machine learning against traditional stochastic models. 
\\
Practical-methodical studies on machine learning trading algorithms occasionally provide comparative data within their specific scope of parameters (for instance, in the case of \cite{khadjeh}, this is given by text mining algorithms with news sentiment inputs).
\clearpage
\section{Research Methodology}

\subsection{Meta-Analysis}

We conduct our investigation using meta-analysis techniques. \cite{Glass} define meta-analysis as the statistical analysis of a large collection of results from individual studies for the purpose of integrating the findings. A similar definition was proposed by \cite{Rose} who state that meta-analysis is a set of quantitative techniques for evaluating and combining empirical results from different studies. Originally designed for application in health sciences, marketing or education \cite{Daniskova}, this technique is increasingly applied in economics and finance where meta-analysis is commonly referred to as meta-regression analysis \cite{Stanley, Jarrell, Stanley2}. Due to the heterogeneity of the subgroups within our sample (i.e., individual experiments conducted by studies), a parametric approach which makes hypotheses based on the comparison of subgroup parameters is unfeasible. The same is true for trying to find factors which influence performance: The lack of standardized testing metrics, standard testing datasets as well as study-specific information on optimization algorithms and weight initialization makes it impossible to form a meaningful meta-regression analysis. These aspects are further detailed in the next subsection. 
\\
Instead, we pursue an approach which evaluates algorithm classes based on their relative rank in subgroup experiments. While this methodology still lacks exhaustive explanatory power on an aggregate level, a pairwise rank analysis based on the same scoring system uncovers meaningful performance differences between algorithm classes.

\subsection{Meta Statistics}
Our data collection procedure encompassed an initial, unfiltered collection of 260 papers. The papers were originally sourced from Google Scholar and SciVerse Science Direct. For each of these sources, we selected the first 50 most relevant papers listed under the key terms "Artificial Intelligence + Financial forecasting", "Machine learning + trading" and "Market prediction + artificial intelligence". Subsequently, we gathered relevant references from these results, added them to the collection and removed duplicates, a procedure which was completed in August 2018. 
Thereafter, we filtered out scientific papers which did not comply with our self-imposed guidelines:
\vspace{1.5mm}
\begin{enumerate}
\item The paper/report demonstrates an application of a machine learning algorithm to forecasting or supporting trading decisions given a time series based on the prices of a publicly traded asset
\item The paper/report provides adequate numerical performance results
\item The paper/report has been published in a peer-reviewed journal or at a peer-reviewed conference
\end{enumerate}
\vspace{1.5mm}
This procedure left us with a total of 170 papers to include in our analysis. From these papers, we extract a total of 2085 performance values from 225 individual experiments (one experiment for every distinct asset with more than one algorithm tested) which we use for the subsequent rank analyses.

\subsection{Dataset}

\subsubsection{Assets}
The studies presented in our dataset encompass an aggregate total of 11 distinct asset classes (stock, index, FX, ETF, mutual fund, commodity, future, option, crypto, bond, money market instrument). In the table of studies, the asset class is indicated in brackets after the specific asset used. If a study presents multiple assets, they are separated with a vertical bar. Furthermore, if the number of assets for a distinct group (e.g., 'stock') exceeds 3, they are not itemized by name. 
\vspace{2mm}
\subsubsection{Market geographies}
This section analyses the market geographies for the asset classes used in the paper. For FX rates, we indicate the geographies pertinent to both currencies, respectively. For reasons of clarity, we do not itemize geographies exceeding three distinct countries.

\begin{table}[htbp]
  \centering
  \captionsetup{font={small,sc},justification=centering}
  \caption{Markets most frequently analyzed by geography and count}
    \begin{tabular}{lr}
    \hline
    Country    & Count \\
    \hline
    UNITED STATES    & 75 \\
    %\hline
    TAIWAN    & 19 \\
    %\hline
    INDIA    & 12 \\
    %\hline
    JAPAN    & 10 \\
    %\hline
    SOUTH KOREA    & 10 \\
    %\hline
    CHINA    & 9 \\
    %\hline
    BRAZIL    & 6 \\
    %\hline
    TURKEY    & 6 \\
    %\hline
    GERMANY    & 5 \\
    %\hline
    SINGAPORE    & 5 \\
    \hline
    \end{tabular}%
  \label{tab:addlabel}%
\end{table}%
\vspace{2mm}
\subsubsection{Periods}

The Input Data represent the periods of data used in individual studies (includes training/testing datasets), with the timestep frequency indicated in brackets. When different periods were used for different assets, these experiments are contextually grouped using a vertical bar. 
\vspace{2mm}
\subsubsection{Input Proxies/Other Inputs}
The 'Input Proxies/Other Inputs' field indicates the usage of features that are not inherent to the time series used by the paper in question. This includes any added information beyond the values of a time series (or transformations of the same). These inputs are represented according to the following taxonomy:

% Table generated by Excel2LaTeX from sheet 'Sheet2'
\begin{table}[htbp]
  \centering
  \captionsetup{font={small,sc},justification=centering}
  \caption{Taxonomy}
    \begin{tabular}{ll}
    \hline
    Variable & Description\\
    \hline
    MARKET & Market data, i.e., data from other assets' t.s. \\
    %\hline
    TECH  & Technical indicators \\
    %\hline
    FUND  & Fundamental corporate finance metrics \cite{Ou} \\
    %\hline
    MACRO & Macroeconomic data \\
    %\hline
    OTHER\{SPECIFY\} & Various \\
    \hline
    \end{tabular}%
  \label{tab:addlabel}%
\end{table}%
\vspace{3mm}
\subsubsection{Algorithms compared}
Our study presents a syntax for creating a high-level understanding of algorithm structures presented by studies on machine learning in financial market prediction. Given the lack of standardization in that field (especially concerning taxonomy), this notation makes a valuable contribution by depicting complex representations in concise terms.
% Table generated by Excel2LaTeX from sheet 'Sheet2'
\begin{table}[htbp]
  \centering
  \captionsetup{font={small,sc},justification=centering}
  \caption{Syntax}
    \begin{tabular}{l l}
    \hline
    Syntax & Description \\
    \hline
    X-Y   & Feed forward \\
    %\hline
    X\^{}Y  & Ensemble \\
    %\hline
    X\{Y\}  & Attribute \\
    %\hline
    X$<$-Y  & Optimization or selection process \\
    %\hline
    $[$X-Y$]$   & Allows for syntax generalization and representation of\\ &complex relationships \\
    \hline
    \end{tabular}%
  \label{tab:addlabel}%
\end{table}%
\vspace{3mm}
\subsubsection{Result metrics}
The result metrics used in studies on financial forecasting using machine learning can roughly be divided into three main groups: Error-based, Return-based, and Accuracy-based. Within our sample, accuracy proved to be the most popular metric, closely followed by annual return and root mean squared error. These groups have different signaling functions related to algorithm- and financial interface performance, which we present and discuss in section IV. \\ The table of studies occasionally contains cells bearing an asterisk; this signals that the study included more metrics than shown within the table which we do not present for reasons of irrelevance or redundancy. Moreover, there are several samples with double asterisks. These signify extrapolation, i.e., integrating an element into our standardized taxonomy even though the study in question does not specifically name the element or is otherwise lacking in information necessary for a definite classification. For this reason, elements marked with two asterisks should be treated with caution as they are based on subjective assumptions given scarce information. It is important to note here that we solely base our rank analysis on performance metrics, excluding metrics such as computational feasibility. 

\subsection{Rank analysis}
Even though the similarities in metrics used across the studies we reviewed appear to suggest a benchmark comparison between individual papers' results, we refrain from conducting a parametric analysis. Notwithstanding the existence of a sufficient amount of performance results for the same algorithm classes for each geography, we identified key differences between studies during our performance analysis which we believe would render a parametric analysis meaningless:
\vspace{5mm}

\textbf{Experimental conditions}

\begin{itemize}
	\item Differences in performance evaluation and reporting 
	\item Different architectures and different practices in varying architectures
	\item Testing environment and validation practices
	\item Length of training/testing sets
	\item Different asset classes and markets (without providing sufficient alpha return metrics)
\end{itemize}
\vspace{5mm}

\textbf{Result evaluation}

\begin{itemize}
    \item Usage of different performance metrics (see section V)
	\item Different ways of annualizing returns
	\item Widely differing trading strategies
\end{itemize}
\vspace{5mm}

Instead, we seek to establish generalizing conclusions from non-parametric analyses on algorithms presented in individual studies. By using an average-over-all approach, we come up with a single rank score between 0 and 1 for a given algorithm type. Our ranking formula separates instances for each paper based on individual algorithms, assets, and performance metrics. Thus, if an algorithm is tested on two assets using three metrics, we receive two instances of three scores which are compiled and later averaged on all studies.
For each algorithm class, this procedure can be expressed as follows:
\begin{equation}
s_{singular}=\frac{1}{N}\sum_{n=1}^{N}\frac{|R_{n}|-r_{n}}{|R_{n}|-1}\
\end{equation}

Where $N$ represents the total number of experiments, counting one experiment per metric, asset, and study. Moreover, $r_{n}$ equals the ranking spot of an algorithm for an individual experiment where $|R_{n}|$ denotes the number of algorithms benchmarked in that experiment. In the case of multiple usages of an algorithm class within an individual experiment (e.g., 'ANN' and 'ANN\{W\}'), we compute an additional average of all ranking spots. Thus, the scoring system allows for multiple classes to attain the same rank score within the same experiment if they have more than one listing in it. This would, e.g., apply to ranks [3,6], [4,5]. Ranks were computed in ascending- or descending order depending on the performance metric used (i.e., ascending for error metrics and descending for accuracy as well as for the majority of return metrics). The results for the most frequently used algorithm classes can be found in Table IV. While these results can certainly be seen as indicative of the overall strength of an algorithm class per se, a direct comparison between classes is not always possible. One algorithm might receive a score which is overall higher than that of another although the two algorithms are never directly compared in an experiment. As a consequence, we ran a pairwise rank analysis visualized in Fig. 1 to be able to directly compare performance between algorithm classes, where 
\vspace{2mm}
\begin{equation}
s_{pairwise}^{(a,b)}=\frac{|\{(y^{(a)},y^{(b)}) \colon y \in Y, y^{(a)}>y^{(b)}\}|}{|\{(y^{(a)},y^{(b)}) \colon y \in Y, y^{(a)}\neq y^{(b)}\}|} 
\end{equation}
\vspace{1mm}
\begin{equation}
Y=Y^{(a)}\frown Y^{(b)}
\end{equation}
\vspace{1mm}
\begin{equation}
Y^{(i)}=\{\frac{|R_{n}^{(i)}|-r_{n}^{(i)}}{|R_{n}^{(i)}|-1}\colon n \in N\} 
\end{equation}
\vspace{2mm}\\
The pairwise rank is computed by performing a simple percentage comparison of two algorithms' relative ranks for individual experiments, $Y^{(a)}$ and $Y^{(b)}$, given that the two algorithms are benchmarked against each other. 
 Fig. 1 displays these pairwise rank scores leading by columns (i.e., the third cell in the first column can be interpreted as evidence that ANNs only perform better than SVMs in 34\% of all surveyed experiments). 'No Data' fields indicate pairs which weren't tested together in any study or bear the same rank scores in all joint experiments
For the purpose of assessing statistical significance, we also conduct a t-test against the null hypothesis that $mean(Y^{(a)})=mean(Y^{(b)})$.
\vspace{2mm}
\section{Results \& Discussion}
\subsection{Rank analysis}
The pairwise rank analysis (see Fig. 1) shows the percentage of times that an algorithm in the column title outperformed its row counterpart. Many of the fields remain empty due to missing data, pointing towards the tendency of studies to compare similar algorithms (e.g., different classes of ANNs), presumably due to the amount of effort involved in constructing fundamentally different model classes. Nevertheless, the pairwise perspective coins several interesting findings. \\ Importantly, given the methodology governing rank scoring and significance tests, observing the sample size in cases where the pairwise rank score is close to 50\% is vital as this may still imply that two algorithm classes perform similarly even though there is no clear winner.
\\
Evidently, the only trading strategy (Buy-and-hold) included in the matrix performs poorly against neural networks and largely does not outperform other algorithms in any scenario. While Buy-and-hold outperforms linear regression models in 32\% of all cases, and random walk in 60\% of all experiments, the differences in rank scores turn out not to be significant at the 5\% significance level. The same holds true for the surprisingly good result against recurrent neural networks which is merely based on two experiments from one study. 
\\
As expected, random walk similarly gets outperformed by ANNs in the vast majority of all experiments. It also scores poorly against AR and GARCH models, and fares surprisingly well against linear regression models, albeit insignificantly so. Finding a clear winner among the traditional statistical models in direct comparison is an arduous task which can largely be explained by the fact that in our sample of studies, these models are most commonly used as a 'traditional' benchmark against various machine learning classes and are rarely tested against each other. Taken from all significant results of statistical models, GARCH models fare best against ANNs. ARIMA score even higher, and though the result is not significant, the large sample size ($>$25) does indicate that the overall performance of ARIMA vs. ANN tends to be more similar than that of GARCH vs. ANN which may suggest that the use of neural networks in returns/price forecasting adds comparatively less value than it does in volatility forecasting.
\\
Interestingly, GARCH models outscore SVMs and appear to fare moderately well against recurrent ANNs (albeit the result is not significant, stemming most likely from a small sample size). A similar pattern can be observed for the pairwise analysis of ANNs and Fuzzy Logic which are frequently used together, thus resulting in closer or equal rank scores per study. \\
It is worthwhile to take a closer look at recurrent neural networks (ANN\{R\}) which significantly outperform other neural networks in our sample. While we do not explicitly list them in the pairwise ranking table due to the limited number of experiments, more recent techniques, such as Long Short-Term Memory ($s_{singular}=0.843$) and Gated Recurrent Unit ($s_{singular}=0.833$) appear to outclass simpler forms of recurrent neural networks which do not explicitly address the vanishing gradient problem, for instance, Elman Networks \cite{Elman} ($s_{singular}=0.580$) although the classes are never directly benchmarked against each other in our sample. Meanwhile, SVMs significantly outscore ANNs which cover similar objectives in classification. While it is difficult to pinpoint the advantages of each method, the significant outperformance of recurrent ANNs against SVMs and other NNs may indicate the relevance of classifiers being able to detect latent temporal patterns in data. 

\begin{table}[htbp]
  \centering
  \captionsetup{font={small,sc},justification=centering}
  \caption{Rank score results for different algorithm classes}
    \begin{tabular}{l l}
    \hline
    Algorithm   & Score \\
    \hline
    SVM   & 0.672 \\
    %\hline
    ANN\{R\} & 0.643 \\
    %\hline
    ANN   & 0.579 \\
    %\hline
    FUZZ  & 0.528 \\
    %\hline
    GARCH & 0.508 \\
    %\hline
    ARIMA & 0.471 \\
    %\hline
    RW    & 0.333 \\
    %\hline
    LRM   & 0.298 \\
    %\hline
    AR    & 0.227 \\
    %\hline
    BH    & 0.167\\
    \hline
    \end{tabular}%
  \label{tab:addlabel}%
\end{table}%

\begin{figure*}[ht]
\includegraphics[width=\textwidth,height=\textheight,keepaspectratio]{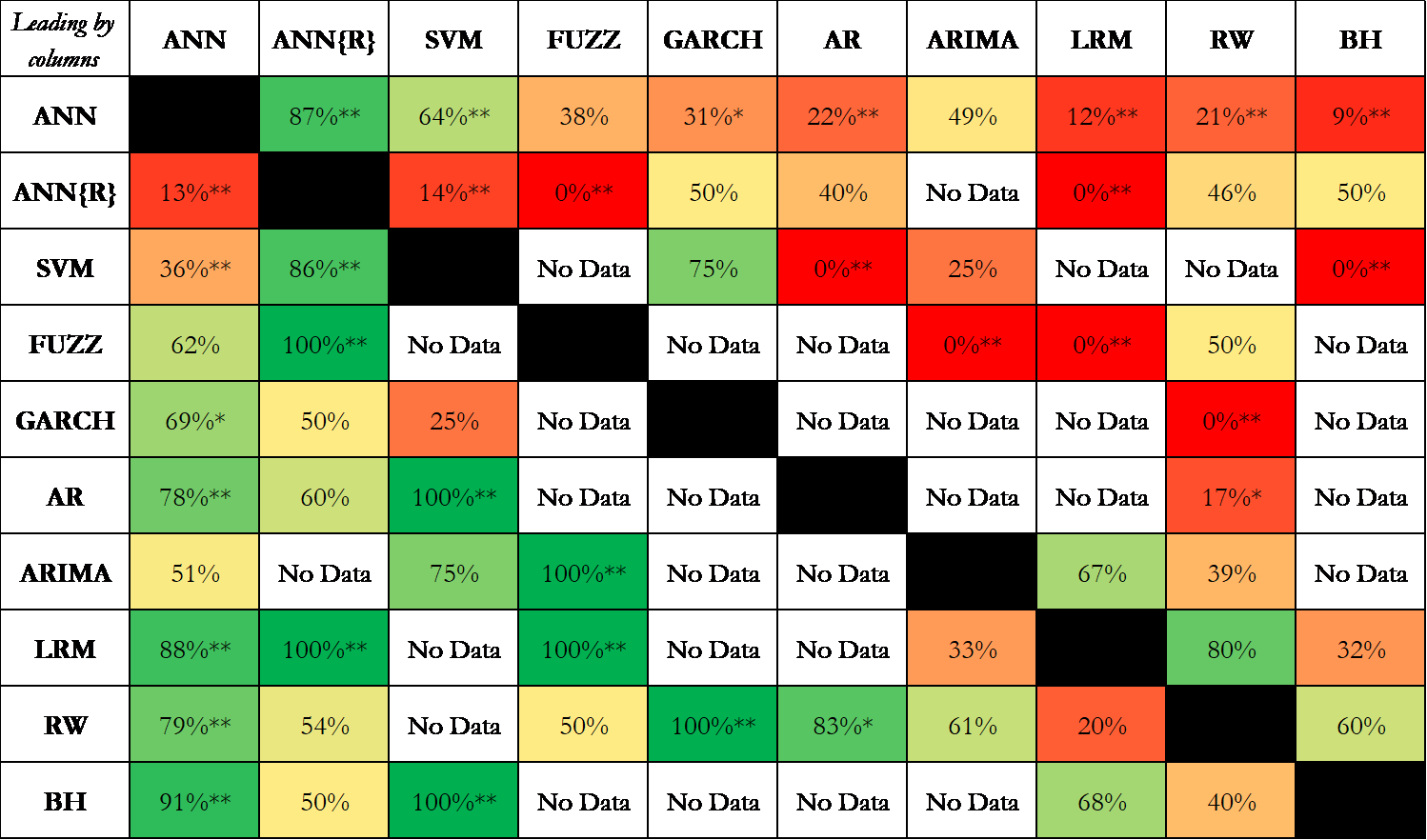}
\captionsetup{font={small,sc}}
\caption{Pairwise rank matrix\\ * for $p\leq0.05$ \\ ** for $p\leq0.01$}
\label{fig}
\end{figure*}

\subsection{Performance metrics for machine learning algorithms in finance}
Relying on accuracy as a performance metric in benchmarking soft computing algorithms in financial applications is problematic. In the papers analyzed within the scope of this meta-analysis, accuracy is most often used in a directional sense. A correct forecast by an algorithm is determined by whether the forecasted variable actually moves in the same direction as the forecast. This definition creates a lack of clarity as some studies define more or less prediction states than others. While most authors limit themselves to forecasting 'Up' or 'Down' movements, others, e.g., \cite{Tsaih} provide three desired output states, making it significantly harder to attain a similar success rate to examples with fewer states. Apart from confusing uninformed readers, this might also hinder direct analyses between different studies. 
Beyond definition issues, it also remains pivotal to be aware of the amount of information on the actual profitability of an algorithm that is carried by the accuracy metric. While accuracy might be a good approximation of an algorithm's general ability, it technically does not convey any information on profitability. Taking an extreme example, an algorithm with high accuracy might correctly forecast many comparably insignificant profit opportunities while missing a small number of large profit opportunities. 
Based on the studies reviewed in this large-scale meta-analysis, we instead advocate in favor of performance metrics which demonstrate the return capabilities of algorithms respective to their fields of forecasting or classification. Relative return metrics, in this context, take into account the magnitude of the trends that a system discovers. One of the most popular return metrics in the meta-analysis proves to be the demonstration of relative outperformance of a reference index (e.g., the S\&P 500).
\\
Going a step further, we propose a method based on \cite{Huang2, Fernandez-Rodriguez, Wah} which takes an ideal classifier system that conducts a trading simulation subject to a pre-defined rule environment/trading strategy, and generates a maximum return indicator by taking optimal (i.e., return-maximizing actions) under all circumstances. The metric itself would then simply form a ratio of an experiment's performance and the ideal classifier.
The rationale behind this metric is aligned with one of the main paradigms of stock forecasting using neural networks itself: Good forecasts are forecasts that generate returns. While metrics such as alpha do capture this logic to some extent, the ratio shown above allows for the definition of more sophisticated trading strategies than 'buy and hold'.
\\
Yet, this metric alone does not exhaustively cover all information needed. For instance, consider that if a system shows low accuracy and high relative return, one could argue that this is the product of learning 'lucky' shared outliers present in both the training and testing set, which is also why our preference does not make the accuracy metric itself redundant. The same logic applies to error metrics: By being application-agnostic, they add valuable information about the viability of the tested methodology regardless of the type of testing data. Accuracy does not capture this; a set of forecasts may exhibit high directional accuracy and high errors at the same time if these forecasts systematically under- or overshoot the true value. This becomes problematic should the underlying system be tested on different data where directional accuracy is less consequential as a performance metric. 
\vspace{2mm}
\section{Conclusion}
In this study, we presented and analyzed a vast array of literature on machine learning applications for financial time series analysis. We collected over 150 relevant papers, forming a large sample containing experiments with different algorithms and asset classes. Following the aim of drawing robust conclusions on the comparative performance of different algorithm classes, we rejected a parametric approach due to the heterogeneity of our literature sample. Instead, we performed purely ranking-based analyses on the performance statistics collected from individual studies, consisting of an aggregate ranking score and a pairwise rank analysis. Our results show significant evidence for the systematic outperformance of machine learning algorithms vs. stochastic models, confirming our initial hypothesis that machine learning algorithms are able to capture meaningful non-linear dynamics in financial time series, and that these dynamics' existence is generalizable across different market geographies and asset class prices. We also demonstrate that recurrent machine learning algorithms tend to perform better at the task of financial market prediction than simple feed forward models, presumably due to their ability to take into account temporal dynamics.
\\
Naturally, these findings have to be put into an appropriate context given the nature of prevailing research. First of all, there is no standardized dataset for machine learning algorithms in financial applications, as opposed to other popular application fields such as image recognition where the MNIST/CIFAR datasets have become a widely accepted standard. Without norms regarding input data, extrapolation based on the performance of an algorithm for one market or one specific asset is impossible, which is why we refrain from a parametric comparison between studies. The lack of standardized input may also exacerbate researcher's bias arising from the desire to achieve a market-beating performance. Given that many machine learning algorithms exhibit a significant black-box characteristic and are highly sensitive to small changes in parameters, they are prone to data manipulation. As a consequence, we identify a strong need for standardized training and testing procedures which will, as a side-effect, also bolster comparability.
\\
Possible steps following this study include collecting a larger amount of studies which specifically test two or more groups of algorithms, i.e., feed forward NNs vs. recurrent NNs or ANNs vs. SVMs. This would be especially interesting for the purpose of comparing sub-classes, such as the ANN\{R\} variants shortly referred to in section V. \\
While the central aim of this meta-analysis is certainly an informative one, we also tried to discover and explain the relationship between the use of certain performance metrics and prevailing biases across studies as well as offering solutions to the same. Ideally, machine learning approaches should be tested on standardized datasets. Alternatively, they should be benchmarked against an ideal classifier to provide a relative perspective on performance. Furthermore, studies should include indications of the algorithm's performance (error metrics such as RMSE) while also relating its performance to a financial interface/trading system (via accuracy- and return metrics). 

\section*{Acknowledgment}

We would like to thank the reviewers from WHU-Otto Beisheim School of Management for their invaluable comments and thoughts. We further extend our expressions of gratitude to Derek Snow from the University of Auckland and all reviewers of the University of Cambridge for their meticulous efforts in correction and revision.

\clearpage
\onecolumn
\newcolumntype{L}{>{\raggedright\arraybackslash}X}
\begingroup
    \fontsize{7pt}{12pt}\selectfont
    \begin{tabularx}{\textwidth}{|>{\RaggedRight}p{2cm}|L|>{\RaggedRight}p{1cm}|L|L|>{\RaggedRight}p{4.6cm}|L|}
    \captionsetup{font={sc}}
    \caption*{Table V: Table of Studies \\ \vspace{4mm}} \tabularnewline
    \hline
    \textbf{Authors, Year} & \textbf{Assets} & \textbf{Market Geographies} & \textbf{Periods} & \textbf{Input Proxies/Other Inputs} & \textbf{Algorithms compared} & \textbf{Result metrics} \\ \hline
    Abraham, Nath \& Mahanti, 2001 & NASDAQ-100 (index) $|$ 6 (stock) $\in$ NASDAQ-100 & US    & 1999-2001 &  & PCA-ANN-NFUZZ\{EFNN\} & RMSE, A \\ \hline
    Adhikari \& Agrawal, 2014 & USD/INR (FX) $|$ GBP/USD (FX) $|$ S\&P 500 (index) $|$ IBM (stock) & US, IN, UK   & 2009-2011 (d) $|$ 1980-1993 (w) $|$ 2004-2007 (d) $|$ 1965-2011 (m) &  & RW-ANN\^{}ANN\{R\{EL\}\}, ANN, ANN\{R\{EL\}\}, RW & MAE, MSE, SMAPE \\ \hline
    Andreou, Neocleous, Schizas \& Toumpouris, 2000 & CSE (index), 10 (stock) $\in$ CSE  & CY    & 1999 & FUND, MACRO, OTHER\{Int. Politics\}, TECH & ANN   & CC, MaxAE, A \\ \hline
    Armano, Marchesi \& Murru, 2005 & COMIT $|$ S\&P 500 (index) & IT, US & 9y (d) & TECH  & NXCS, BH & AR\%, std, SR, A\{HR\}, * \\ \hline
    Atiya, Talaat \& Shaheen, 1997 & S\&P 500 (stock) & US    & 1993-1994** & FUND  & ANN, BH & AR \\ \hline
    Bagheri, Peyhani \& Akbari, 2014 & EUR/USD $|$ USD/JPY $|$ GBP/USD $|$ USD/CHF (FX) & Various & 2011-2014 (d) & TECH  & NFUZZ\{ANFIS\}, FUZZ\{M\}, FUZZ\{TSK\}, NFUZZ & A\{HR\} \\ \hline
    Banik, Chanchary, Rouf \& Khan, 2007 & DSPI (index) & BGD   & 2003-2007 (d) &  & NFUZZ\{ANFIS\}, ANN, ARIMA & MAE, MAPE, RMSE, RMSPE, R2 \\ \hline
    Bildirici \& Ersin, 2009 & ISE 100 (index) & TR    & 1987-2008 (d) &  & GARCH, GARCH\{E\}, GARCH\{T\}, GARCH\{GJR\}, GARCH\{SA\}, GARCH\{POW\}, GARCH\{N\}, GARCH\{AP\}, GARCH\{NP\}, ANN\^{}GARCH, ANN\^{}GARCH\{E\}, ANN\^{}GARCH\{T\}, ANN\^{}GARCH\{GJR\}, ANN\^{}GARCH\{SA\}, ANN\^{}GARCH\{POW\}, ANN\^{}GARCH\{N\}, ANN\^{}GARCH\{AP\}, ANN\^{}GARCH\{NP\} & RMSE \\ \hline
    Bildirici \& Ersin, 2013 & USD/bWTI (commodity) & US    & 1986-2012 (d) &  & GARCH, GARCH\{AP\}, GARCH\{FI\}, GARCH\{FIAP\}, LSTAR-LST-GARCH, LSTAR-LST-GARCH\{AP\}, LSTAR-LST-GARCH\{FI\}, LSTAR-LST-GARCH\{FIAP\}, ANN-GARCH, ANN-GARCH\{AP\}, ANN-GARCH\{FI\}, ANN-GARCH\{FIAP\}, LSTAR-LST-ANN-GARCH, LSTAR-LST-ANN-GARCH\{AP\}, LSTAR-LST-ANN-GARCH\{FI\}, LSTAR-LST-ANN-GARCH\{FIAP\},  & RMSE, PC \\ \hline
    Bildirici, Alp \& Ersin, 2010 & ISE 100 (index) $|$ TRY/USD & TR, US & 1987-2008 (m) & MARKET & TAR-VEC, TAR-VEC$<$Hansen \& Seo, 2002$>$, ANN\{TAR-VEC\}, ANN\{RBF\{TAR-VEC\}\}, ANN\{HE\{TAR-VEC\}\} & RMSE \\ \hline
    Bodyanskiy \& Popov, 2006 & DJIA (index) & US    &  &  & GARCH, ANN\{RMD\}, ANN\{QP\} & NMSE, NMAE, A\{HR\}, A\{WHR\} \\ \hline
    Cao \& Tay, 2001 & S\&P 500 (index) & US    & 1993-1995 (d) & TECH  & ANN, SVM & NMSE, MAE, DS, CP, CD \\ \hline
    Cao \& Tay, 2003 & 5 (future) $\in$ CMM & US   & 1988-1999, * & TECH  & SVM, ANN, ANN\{RBF\}, SVM\{A\}, ANN\{WBP\} & NMSE, MAE, DS \\ \hline
    Cao, Leggio \& Schniederjans, 2005 & 367 (stock) $\in$ SHSE & CN    & 1999-2002 (d) & FUND  & LRM\{uni\}, ANN\{uni\}, LRM\{multi\}, ANN\{multi\} & MAE, MAPE, MSE \\ \hline
    Casas, 2001 & x (stock) (bond) (MM) & US    & 1994-1999 & MARKET, MACRO & ANN   & AR \\ \hline
    Chang \& Liu, 2008 & 1 (stock) $\in$ TSE $|$ TSE (index) & TW    & 2002-2006 (d) $|$ 2003-2005 (d) & TECH  & FUZZ\{TSK\}, ANN, LRM\{multi\} & MAPE \\ \hline
    Chang, Fan \& Liu, 2009 & 9 (stock) & CN    & 2004-2006 & TECH  & PLR$<$-GA-ANN, ANN, PLR  & AR, A\{HR\} \\ \hline
    Chang, Liu, Lin, Fan \& Ng, 2009 & 9 (stock) & TW    & 2004-2005 (d) & TECH  & ANN\{CBR\}, CBR, ANN & AR \\ \hline
    Chang, Wang \& Zhou, 2012 & 1 (stock) $\in$ TSE $|$ TSE (index) & TW    & 2003-2006 (d) $|$ 2005 (d) & TECH  & ANN\{PC\}$<$-GA, FUZZ\{TSK\}, ANN, LRM & MAPE \\ \hline
    Chavarnakul \& Enke, 2008 & S\&P 500 (index) & US    & 1998-2003 (d) & TECH  & ANN\{GR\} & MSE, SIGN \\ \hline
    Chen \& Leung, 2004 & GBP/USD $|$ CAD/USD $|$ JPY/USD (FX) & Various & 1980-2001 (m) & MACRO & ANN\{GR\}, MTF, GMM, BVAR, MTF-ANN\{GR\}, GMM-ANN\{GR\}, BVAR-ANN\{GR\}, RW & AR, RMSE, ThU \\ \hline
    Chen, 1994 & 4** (stock) &  & 1986-1992 & TECH  & AR, ANN, ANN\{GR\}, ANN\{CS\} & CC, RMSE \\ \hline
    Chen, Abraham, J. Yang \& B. Yang, 2005 & NASDAQ-100 $|$ NIFTY (index) & US, IN & 1995-2002 $|$ 1998-2001 &  & ANN, FUZZ\{T-S\}, FUZZ\{H-T-S\} & RMSE, MAP, MAPE, CC \\ \hline
    Chen, Dong \& Zhao, 2005 & NASDAQ-100 $|$ NIFTY (index) & US, IN & 1995-2002 $|$ 1998-2001 &  & ANN\{LLWAV\}, ANN\{WAV\} & CC, MAP, MAPE, RMSE \\ \hline
    Chen, Leung \& Daouk, 2003 & TAIEX** (index) & TW    & 1982-1992 & MACRO & ANN\{P\}, KF, RW, BH & AR, * \\ \hline
    Chen, Ohkawa, Mabu, Shimada \& Hirasawa, 2009 & 10 (stock) $\in$ TSM & JP    & 2001-2004 (d) & TECH  & GNP\{CN\}, GNP\{RL\}, GNP\{Candlestick\}, GA, BH & AR \\ \hline
    Chen, Shih \& Wu, 2006 & Nikkei 225 $|$ All Ordinaries $|$ Hang Seng $|$ Straits Times $|$ TAIEX $|$ KOSPI $|$ (index) & Various & 1971-2002 (d) & TECH  & SVM, ANN, AR & MSE, NMSE, MAE, DS, WDS \\ \hline
    Chen, Yang \& Abraham, 2007 & NASDAQ-100 $|$ NIFTY (index) & US, IN & 7y $|$ 4y (d) &  & DT\{FNT\}, ENS[DT]\{B\}, ENS[DT]\{G\}, ENS[DT]\{LWPR\} & RMSE, MAP, MAPE, CC \\ \hline
    Chenoweth \& Obradovic, 1996 & S\&P 500 (index) & US    & 1985-1993 (d) & MACRO & ANN, ANN\^{}ANN & AR, BETC \\ \hline
    Chenoweth, Obradovic \& Lee, 1996 & S\&P 500 (index) & US    & 1982-1993 (d) & MARKET, TECH & ANN, BH & AR, BETC \\ \hline
    Chiang, Urban \& Baldridge, 1996 & 101 (mutual fund) & US    & 1981-1986 (y) & MACRO & ANN, LRM, NLRM & MAPE, * \\ \hline
    Chong, Han \& Park, 2017 & 38 (stock) $\in$ KOSPI  & KR    & 2010-2014 (5-min) &  & AR, ANN, ANN\{D\}, AE-ANN\{D\}, PCA-ANN\{D\}, RBM-ANN\{D\} & NMSE, RMSE, MAE, MI \\ \hline
    Chun \& Park, 2005 & KOSPI (index) & KR    & 2000-2004 (d) &  & ENS\{CBR\{DA\}\}, ENS\{CBR\{SE\}\}, RW & MAPE \\ \hline
    Constantinou, Georgiades, Kazandjian \& Kouretas, 2006 & CSE (index) & CY    & 1996-2002 (d) &  & ANN, MarkS & RMSE \\ \hline
    Dai, Wu \& Lu, 2012 & Nikkei 225 $|$ Shanghai B-Share stock index (index) & JP, CN & 2004-2009 (d) & MARKET & NLICA-ANN, LICA-ANN, PCA-ANN, ANN & RMSE, MAE, MAPE, RMSPE, DS \\ \hline
    Das, Mishra \& Rout, 2017 & USD/INR $|$ USD/EUR (FX) & US, IN, EU & 2001-2016 (d) & TECH  & ELM, ANN\{FL\}, ANN & MSE, MAPE, MAE, ThU, ARV \\ \hline
    de A. Araujo, Nedjah, M. de Seixas, L.I. Oliveira \& R. de L. Meira, 2018 & 5 (stock) $\in$ BOVESPA & BR    &  &  & ARIMA, ANN, ANN\{RBF\}, SVR\{L\}, SVR\{POLY\}, SVR\{RBF\}, IDL$<$-BP, IDL$<$-GA, IDL$<$-PSO, IDL$<$-BSA, IDL$<$-FFA, IDL$<$-CS & ARV, MAPE, MSE, POCID, ThU, EF \\ \hline
    de C. T. Raposo \& de O. Cruz, 2002 & 28 (stock) $\in$ SPSE & BR    & 1986-1998 & FUND  & PCA-NFUZZ & A \\ \hline
    de Faria, Marcelo Albuquerque, Gonzalez, Cavalcante \& Marcio Albuquerque, 2009 & Bovespa (index) & BR    & 1998-2008 (d) &  & ANN, AES & RMSE, A \\ \hline
    de Oliveira, Nobre \& Zarate, 2013 & 1 (stock) $\in$ BM\&FBOVESPA  & BR    & 2000-2011 (m) & MARKET, MACRO, TECH & ANN   & MAPE, RMSE, ThU, POCID \\ \hline
    Dempster \& Leemans, 2006 & EUR/USD (FX) & EU, US & 2000-2002 (1-min) &  & RL\{R\} & AR \\ \hline
    Doeksen, Abraham, Thomas \& Paprzycki, 2005 & 2 (stock) & US    & 1997-2003 (d) & MACRO, TECH & ANN, FUZZ\{M\}$<$-GA, FUZZ\{TSK\}$<$-GA & MSE, AR, A \\ \hline
    Dunis, Laws \& Sermpinis, 2011 & EUR/USD (FX) & EU, US & 1994-2001 (d) & MARKET, MACRO & ANN\{PS\}, ANN\{HO\}, ANN\{R\}, ANN, ANN\{SCE\}, ANN\{GM\}, MACDM, ARMA, LOGIT, NAIVE & SR, AR, MD, VOLA, * \\ \hline
    Enke \& Thawornwong, 2005 & S\&P 500 (index) & US    & 1976-1999 (m) & FUND, MACRO, MARKET & ANN, ANN\{GR\},  ANN\{P\}, LRM, BH & CC, RMSE, A, AR, std, SR, * \\ \hline
    Fatima \& Hussain, 2008 & KSE100 (index) & PAK   & 2000-2002 (d) &  & ARIMA, GARCH, ANN, ARIMA-ANN, GARCH-ANN & FMSE \\ \hline
    Fernandez \& Gomez, 2007 & 5 (index) & Various & 1992-1997 (w) & MARKET, TECH** & ANN\{H\}, GA, TS, SimA & MPE \\ \hline
    Fernandez-Rodriguez, Gonzalez-Martel \& Sosvilla-Rivero, 2000 & IGBM (index) & ESP   & 1966-1997 (d) &  & ANN   & A, alpha\%, IPR, SR \\ \hline
    Fischer \& Krauss, 2018 & S\&P 500 (stock) & US    & 1990-2015 (d) &  & ANN\{R\{LSTM\}\}, RF, ANN, LRM & AR, SR, SortR, MD, * \\ \hline
    Freitas, de Souza \& de Almeida, 2009 & 52 (stock) $\in$ Bovespa & BR    & 1999-2007 (w) &  & ANN\{ARMR\} & ME, RMSE, A, MAPE, AR \\ \hline
    Ghazali, Hussain, Nawi \& Mohamad, 2009 & 4 (FX) & Various & 2000-2005 (d) & TECH  & ANN, ANN\{PS\}, ANN\{RP\}, ANN\{DRP\}, RW, LRM, ARIMA & AR, VOLA, NMSE, A, RM, MAPE \\ \hline
    Grudnitski \& Osburn, 1993 & S\&P 500 (index) $|$ (future) -$>$ Gold & US    & 1982-1990 (m) & MACRO, MARKET, TECH & ANN   & alpha, A \\ \hline
    Guresen \& Kayakutlu, 2008 & XU100 (index) & TR    & 2003-2008 (d) &  & ANN, LTS, ANN\{R\}, ANN\{DAN2\}, GARCH-ANN, GARCH\{E\}-ANN, GARCH-LTS, GARCH\{E\}-LTS, GARCH-ANN\{R\}, GARCH\{E\}-ANN\{R\}, GARCH-ANN\{DAN2\}, GARCH\{E\}-ANN\{DAN2\} & MSE, MAE, MAPE \\ \hline
    Guresen, Kayakutlu \& Daim, 2011 & NASDAQ (index) & US    & 2008-2009 (d) &  & ANN, ANN\{DAN2\}, GARCH-ANN, GARCH-ANN\{DAN2\} & MSE, MAE, MAPE \\ \hline
    Hajizadeh, Seifi, Zarandi \& Turksen, 2012 & S\&P 500 (index) & US    & 1998-2009 (d) & MARKET, MACRO, TECH & GARCH\{E\}-ANN, GARCH, GARCH\{E\}, GARCH\{GJR\} & RMSE, MAE, MAPE, MFE \\ \hline
    Harvey, Travers \& Costa, 2000 & Emerging market indices \& composites & Various & 1997-1999 (w) & FUND, MARKET & ANN, BH & AR, DA, MM \\ \hline
    Hassan, 2009 & BAY $|$ DAL $|$ RYA $|$ AAPL $|$ IBM $|$ DELL (stock) & US & 2002-2004 (d, do, dh, dl) &  & HMM-FUZZ, ARIMA, ANN & MAPE \\ \hline
    Hassan, Nath \& Kirely, 2007 & AAPL $|$ IBM $|$ DELL (stock) & US    & 2003-2005 (d, do, dh, dl) &  & HMM, [ANN-HMM]$<$-GA, [ANN-HMM]$<$-GA-WA, ARIMA & MAPE \\ \hline
    Hsieh, Hsiao \& Yeh, 2011 & DJIA $|$ FTSE $|$ Nikkei 225 $|$ TAIEX (index) & Various & 1997-2003 \& 2002-2008 & TECH  & ANN\{R\{W\}\}, ANN, ANN$<$-ABC,  FUZZ$<$Chen$>$, FUZZ$<$Yu$>$, NFUZZ\{ANFIS\} & RMSE, MAE, MAPE, AR \\ \hline
    Huang \& Tsai, 2009 & FITX (index) & TW    & 2000-2006 (d) & TECH  & SOM-SVR, SVR & MSE, MAE, MAPE \\ \hline
    Huang \& Wu, 2008 & 7 (index) & Various & 2003-2005 & MARKET & GA-SVM, GARCH, ANN, SVM & RMSE \\ \hline
    Huang, Nakamori \& Wang, 2005 & Nikkei 225 (index) & JP    & 1990-2002 (w) & MACRO, MARKET & RW, LDA, QDA, ANN\{R\{EL\}\}, SVM, RW\^{}LDA\^{}QDA\^{}ANN\{R\{EL\}\}\^{}SVM  & A\{HR\} \\ \hline
    Huang, Pasquier \& Quek, 2009 & HSI (index) $|$ 1 (stock)  & HK, SGP & 1987-2006 (d), * & TECH  & NFUZZ\{EFNN\}, NFUZZ\{DENFIS\}, NFUZZ\{RSPOP\}, NFUZZ\{HiCEFS\}, BH & AR \\ \hline
    Huarng \& Yu, 2008 & TAIEX (index) & TW    & 1999-2004 (d) & MARKET & FUZZ\{uni\}$<$Chen, 1996$>$, LRM\{uni\}, ANN\{uni\}, ANN-FUZZ\{uni\}, ANN-FUZZ\{subs\}\{uni\}, LRM\{bi\}, ANN\{bi\}, ANN-FUZZ\{bi\}, ANN-FUZZ\{subs\}\{bi\} & RMSE \\ \hline
    Hussain, Knowles, Lisboa \& El-Deredy, 2008 & EUR/USD $|$ JPY/USD $|$ GBP/USD (FX) & Various & 1994-2001 (d) &  & ANN\{PP\}, ANN\{FL\}, ANN & AR, MD, MSE \\ \hline
    J.-Z. Wang, J.-J. Wang, Zhang \& Guo, 2011 & SCI (index) & CN    & 1993-2009 (m) &  & ANN\{WAV\}, ANN & MAE, RMSE, MAPE \\ \hline
    Kanas \& Yannopoulos, 2001 & FTAI $|$ DJIA (index) & UK, US & 1980-2000 (m) & FUND, TECH & ANN, LRM & RMSE \\ \hline
    Kara, Boyacioglu \& Baykan, 2011 & ISE 100 (index) & TR    & 1997-2007 (d) & TECH  & ANN, SVM, OLS, ANN$<$Diler, 2003$>$, ANN$<$Altay \& Satman, 2005$>$ & A \\ \hline
    Khan, 2011 & Nikkei 225 (index) & JP    & 1996-2009 (5-min) &  & HAR, SVM-HAR, HAR\{J\}, HAR\{MSNR, J\}, SVM-HAR\{J\}, SVM-HAR\{MSNR, J\} & RMSE, MAE, RMSPE, MAPE \\ \hline
    Khemchandani, Jayadeva \& Chandra, 2009 & 5 (stock) $|$ S\&P 500 (index) & US    & 2005-2007, * $|$ 1989-1993 (d) &  & SVR\{RLSF\}, SVR & NMSE \\ \hline
    Kim \& Ahn, 2012 & KOSPI (index) & KR    & 1989-1998 (d) & TECH  & ANN$<$-GA, ANN & A \\ \hline
    Kim \& Chun, 1998 & SGPI (index) & SGP   & 1985-1996 (d) & FUND, TECH & ANN\{P\}, ANN\{R\}, CBR, ANN & A\{HR\} \\ \hline
    Kim \& Han, 2000 & KOSPI (index) & KR    & 1989-1998 (d) & TECH  & GA-ANN$<$-GA, ANN, ANN$<$-GA & A\{HR\} \\ \hline
    Kim \& Shin, 2007 & KOSPI 200 (index) & KR    & 1997-1999 (d) &  & ANN\{ATD\}$<$-GA, ANN\{TD\}$<$-GA, ANN\{ATD\}, ANN\{TD\}, ANN\{R\} & MSE \\ \hline
    Kim, 2003 & KOSPI (index) & KR    & 1989-1998 (d) & TECH  & SVM, ANN, CBR & A\{HR\} \\ \hline
    Kim, 2006 & KOSPI (index) & KR    & 1991-1998 (d) & TECH  & GA\{CBR\}-ANN, ANN & A\{HR\} \\ \hline
    Kim, Han \& Chandler, 1998 & (future) -$>$ KOSPI 200 & KR    & 1996 (d) & TECH  & ANN\^{}CBR, BH & A\{HR\}, AR \\ \hline
    Kimoto, Asakawa, Yoda \& Takeoka, 1990 & TOPIX (index) & JP    & 1985-1989 (w) & MARKET, FUND**, MACRO, TECH** & ENS[ANN] & CC \\ \hline
    Ko \& Lin, 2008 & 21 (stock) $\in$ Taiwan 50 & TW    & 2000-2005 &  & ANN\{RA\} & AR  \\ \hline
    Koulouriotis, Diakoulakis, Emiris \& Zopounidis, 2005 & ASE (index) & GRE   & 1996-1997 (w, *) & MARKET, TECH & LRM, ANN, ANN$<$-GA, ANN\{RBF\}, NFUZZ\{ANFIS\}, ANN\{DC\} & A, MSE \\ \hline
    Krauss, Do \& Huck, 2017 & S\&P 500 (stock) & US    & 1990-2015 (d) &  & ANN, DT\{GB\}, RF, [ANN\^{}DT\{GB\}\^{}RF]\{WA\} & AR, ER, SR, MD, SortR, * \\ \hline
    Kristjanpoller \& Minutolo, 2015 & Gold (commodity) $|$ (future) -$>$ Gold &  & 1999-2014 (d) & MARKET & GARCH-ANN, GARCH & MAPE, MAE, MSD \\ \hline
    Kristjanpoller \& Minutolo, 2016 & Oil (commodity) $|$ (future) -$>$ Oil & US    & 2002-2014 (d) & MARKET & GARCH-ANN, GARCH, ARFIMA & HMSE, HMAE, * \\ \hline
    Kristjanpoller \& Minutolo, 2018 & USD/BTC (crypto) & US    & 2011-2017 (d) & TECH  & ANN-GARCH\{E\}, GARCH\{E\}, * & MSE \\ \hline
    Kristjanpoller, Fadic \& Minutolo, 2014 & Bovespa $|$ IPSA $|$ IPyC (index) & BR, CHI, MEX & 2000-2011 (d) &  & GARCH, GARCH-ANN & MSE, RMSE, MAE, MAPE, MAPE reduction \\ \hline
    Kryzanowski, Galler \& Wright, 1993 & 120 (stock) & CAN** & 1981-1991 & FUND, MACRO & ANN\{BM\} & A, * \\ \hline
    Kumar, Meghwani \& Thakur, 2016 & 12 (index) & Various & 2008-2013 & TECH  & SVM\{PROX\}, LC-SVM\{PROX\}, RC-SVM\{PROX\}, RR-SVM\{PROX\}, RF-SVM\{PROX\}, ANN, LC-ANN, RC-ANN, RR-ANN, RF-ANN & A \\ \hline
    Kuo, 1998 & x (stock) $\in$ TAIEX & TW    &  & TECH, MARKET & FUZZ\^{}ANN-ANN & MSE, AR, * \\ \hline
    Kuo, L. C. Lee \& C. F. Lee, 1996 & TAIEX** (index) & TW    & 281d  & MARKET, TECH & ANN\^{}FUZZ\{Delphi\}-ANN, ANN & MSE, AR, * \\ \hline
    Kwon \& Moon, 2007 & 36 (stock) $\in$ NYSE/NASDAQ & US    & 1992-2004 (d, dh, dl) & TECH  & ANN\{R\{EL\}\}$<$-GA, GA-$>$CBE & Instance-based alpha \\ \hline
    Lam, 2004 & 364 (stock) $\in$ S\&P 500 & US    & 1985-1995 & FUND, MACRO, TECH & ANN   & AR \\ \hline
    Lee \& Chen, 2002 & Nikkei 225 $|$ MSCI Taiwan (index) & JP, TW & 1998-1999 (5-min) & MARKET & ANN, RW, GARCH & RMSE, MAE, MAPE, RMSPE \\ \hline
    Lee \& Chiu, 2002 & Nikkei 225 (index) & JP    & 1998-1999 (5-min) & MARKET & ANN, RW & RMSE, MAE, MAPE, RMSPE \\ \hline
    Lee, 2009 & NASDAQ (index) & US    & 2001-2007 (d) & MARKET & SVM, FSSFS-SVM, IG-SVM, SU-SVM, CFS-SVM, ANN, FSSFS-ANN, IG-ANN, SU-ANN, CFS-ANN & A \\ \hline
    Lee, Cho \& Baek, 2003 & (future) -$>$ KOSPI 200 & KR    & 1999-2001 & TECH  & ANN\{AA\} & MAE \\ \hline
    Leigh, Paz \& Purvis, 2002 & NYSE (index) & US    & 1980-1999 (d) &  & ANN\{CBR\} & A\{HR\} \\ \hline
    Lendasse, de Bodt, Wertz \& Verleysen, 2000 & Bel 20 (index) & BEL   & 10y (d) & TECH, MARKET, MACRO & PCA-CCA-ANN\{RBF\} & A \\ \hline
    Leu, Lee \& Jou, 2009 & NTD/USD (FX) & TW, US & 2006-2007 (d) & MARKET & FUZZ\{DB\}, RW, ANN\{RBF\} & MSE, DS \\ \hline
    Li \& Kuo, 2008 & TAIEX (index) & TW    & 1991-2002 (d) & TECH  & SOM, SOM\{DWT\} & AbR, A\{HR\}, * \\ \hline
    Li, Zhang, Wong \& Qin (2009) & S\&P 500 $|$ FTSE100 $|$ Nikkei 225 (index) $|$ USD/EUR $|$ USD/GBP $|$ USD/JPY (FX) & Various & 2000-2003 (d) &  & RW\^{}AES\^{}ARIMA\^{}ANN-RBF, RW, AES, ARIMA & ER \\ \hline
    Liao \& Wang, 2010 & SAI $|$ SBI $|$ HSI $|$ DJIA $|$ IXIC $|$ S\&P 500 (index) & CN, HK, US & 1990-2008 (d) &  & ANN\{STE\} & ARE \\ \hline
    Lin \& Yeh, 2009 & x (option) $\in$ TAIFEX & TW    & 2003-2004 & MARKET, TECH & ANN, Grey-ANN, GARCH-ANN & MAE, MAPE \\ \hline
    Lu \& Wu, 2011 & Nikkei 225 $|$ TAIEX (index) & JP, TW & 2004-2008 (d) & MARKET & ANN\{CMAC\}, SVR, ANN & RMSE, MAE, MAPE, A, * \\ \hline
    Lu, Que \& Cao, 2016 & Chinese energy index (index) & CN    & 2013-2016 &  & GARCH\{E\}-ANN, GARCH\{GJR\}-ANN, GARCH\{E\}\^{}ANN, GARCH\{GJR\}\^{}ANN & RMSE \\ \hline
    M.-Y. Chen,  D.-R. Chen, Fan \& Huang, 2013  & TAIEX (index) & TW    & 2000-2010 (d) & MARKET & FUZZ\{FTS\{W\}\}, NFUZZ, NFUZZ\{ANFIS\}$<$Cheng, Wei \& Chen, 2009$>$, NFUZZ\{AR-ANFIS\}, ANN\{R\{W\}\}, NFUZZ\{ANFIS\}, ANN & RMSE \\ \hline
    M.-Y. Chen, Fan, Y.-L. Chen \& Wei, 2013 & Taiwan 50 (index) $|$ 40 (stock) $\in$ NYSE & TW, US & 2006-2011 (d), * & FUND  & ANN, ANN\{RBF\}, SVR, DOE-ANN, LRM, LMS & PC, RMSE \\ \hline
    Majhi, G. Panda, Sahoo, A. Panda \& Choubey, 2008 & S\&P 500 $|$ DJIA (index) & US    & 1994-2006 (d) & TECH  & ALC$<$-PSO, ANN & MAPE \\ \hline
    Majhi, Panda \& Sahoo, 2009 & USD/INR $|$ USD/GBP $|$ USD/JPY (FX) & Various & 1971-2005 (m), * & TECH  & LMS, ANN\{FL\}, ANN\{CFL\} & MSE, APE \\ \hline
    Majhi, Panda, Sahoo, Dash \& Das, 2007 & S\&P 500 $|$ DJIA (index) & US    & 1994-2006 (d) & TECH  & ANN, ALC$<$-BFO & MAPE \\ \hline
    Malliaris \& Salchenberger, 1993 & x (option) & US    & 1990 (dm, de) & MARKET, MACRO & ANN, NLM\{Black Scholes\} & MAE, MAPE, MSE \\ \hline
    Mizuno, Kosaka, Yajima \& Komoda, 1998 & TOPIX (index) & JP    & 1982-1987 (w) & TECH  & ANN   & A \\ \hline
    Monfared \& Enke, 2014 & NASDAQ (index) & US    & 1997-2011 & MARKET & GARCH\{GJR\}, GARCH\{GJR\}-ANN, GARCH\{GJR\}-ANN\{GR\}, GARCH\{GJR\}-ANN\{RBF\}  & MSE, MSE reduction \\ \hline
    Motiwalla \& Wahab, 2000 & 11 (index) & US    & 1990-1998 (m) & MARKET, MACRO & BH, ANN, LRM & AR, SR, std, A \\ \hline
    Nayak, Misra \& Behera, 2012 & BSE S\&P 100 $|$ BSE S\&P 500 (index) & IN    & 2005-2010 & MARKET, MACRO & ANN, ANN$<$-GA, ANN\{FL\}$<$-GA & MAE \\ \hline
    Ni \& Yin, 2009 & USD/GBP (FX) & US, UK & 4000 days** (d) & TECH  & SOM-SVR, SOM-ANN, GARCH, SOM\{R\}-SVR-GA & A \\ \hline
    Oh \& Kim, 2002 & KOSPI 200 (index) & KR    & 1990-2000 (d) &  & BH, ANN, ANN\{PWNL\} & RMSE, MAE, MAPE, AR \\ \hline
    Olson \& Mossman, 2003 & x (stock)  & CAN   & 1976-1993 & FUND  & OLS, LOGIT, OLS-ANN, LOGIT-ANN & A\{HR\}, AR \\ \hline
    Pai \& Lin, 2005 & 10 (stock) & US    & 2002-2003 (d) &  & ARIMA, SVM, ARIMA\^{}SVM, ARIMA-SVM & MAE, MSE, MAPE, RMSE \\ \hline
    Pan, Tilakaratne \& Yearwood, 2005 & AORD (index) & AUS   & 1990-2003 & MARKET & ANN   & RMSE, A, VR \\ \hline
    Panda \& Narasimhan, 2007 & INR/USD (FX) & IN, US & 1994-2003 (w) &  & ANN, AR, RW & RMSE, MAE, MAE, PC, DA, SIGN \\ \hline
    Pantazopoulos, Tsoukalas, Bourbakis, Brun \& Houstis, 1998 & S\&P 500 (index) & US    & 1928-1993 (d) & MARKET, TECH & NFUZZ, BH & RMSE, AR \\ \hline
    Perez-Rodriguez, Torra \& Andrada-Felix, 2005 & Ibex 35 (index)  & ESP   & 1989-2000 (d) &  & AR, ANN, ANN\{R\{EL\}\}, STAR\{E\}, LSTAR, AR\^{}ANN & AR, SR, MAE, MAPE, RMSE, ThU, SIGN, DA \\ \hline
    Petropoulos, Chatzis, Siakoulis \& Vlachogiannakis, 2017 & 10 (FX)/USD & Various & 2001-2015 (d) &  & [NBAY\^{}SVM\^{}ANN\^{}RF\^{}BART\^{}N\^{}BH\^{}SH\^{}AR]-GA, [NBAY\^{}SVM\^{}ANN\^{}RF\^{}BART\^{}N\^{}BH\^{}SH\^{}AR]-CLS, [NBAY\^{}SVM\^{}ANN\^{}RF\^{}BART\^{}N\^{}BH\^{}SH\^{}AR]-V\{MAJ\}, [NBAY\^{}SVM\^{}ANN\^{}RF\^{}BART\^{}N\^{}BH\^{}SH\^{}AR]-VAR & AR, SR, VOLA, MD, * \\ \hline
    Qi, 1999 & S\&P 500 (index) & US    & 1954-1992 & FUND, MACRO & LRM, ANN & AR, std, SR, A, RMSE, MAE, MAPE, CC \\ \hline
    Quah, 2008 & 1630 (stock) $\in$ DJIA & US    & 1995-2004 (d, *) & FUND  & ANN, NFUZZ\{ANFIS\}, ANN\{RBF\{GGAP\}\} & A, AR \\ \hline
    Quek, Yow, Cheng \& Tan, 2009 & 23 (stock) $\in$ NASDAQ, NYSE, * & US    & 1996-2005 (d) & TECH  & NFUZZ\{SO\} & AR \\ \hline
    R. Dash \& P. K. Dash, 2016 & BSE SENSEX $|$ S\&P 500 (index) & IN, US & 2010-2014 (d) & TECH  & ANN\{FL\}$<$-ELM, SVM, NBAY, kNN, DT & AR \\ \hline
    Rast, 1999 & DAX (index) & GER   & 1985-1987 \& 1996-1998 (d) &  & ANN, NFUZZ & AR \\ \hline
    Rather, 2011 & 6 (stock) $\in$ NSE & IN    & 2007-2010 (w) &  & ANN\{CBR\} & ME, MSE, MAPE \\ \hline
    Rather, 2014 & 3 (stock) $\in$ BSE & IN    & 2013 (d) &  & AR-ANN\{R\{ARMR\}\} & MSE, MAE \\ \hline
    Rather, Agarwal \& Sastry, 2015 & 25 (stock) $\in$ BSE & IN    & 2007-2010 (w) \& 2013 (d) &  & [ANN\{ARMR\}\^{}ES\^{}ARMA]$<$-GA, ANN\{R\} & MSE, MAE, CC \\ \hline
    Refenes, Azema-Barac \& Zapranis, 1993 & 143 (stock) & UK    & 1985-1991 &  & ANN, LRM & RMSE \\ \hline
    Rodriguez-Gonzalez, Garcia-Crespo, Colomo-Palacios, Guldris Iglesias \& Gomez-Berbis, 2011 & 15 (stock) $\in$ IBEX 35 & ESP   & 16 years (d) & TECH  & ANN\{G\} & A \\ \hline
    Rout, P.K. Dash, R. Dash \& Bisoi, 2017 & BSE $|$ S\&P 500 (index) & IN, US & 2004-2008 (d) $|$ 2010-2012 (d) & TECH  & ANN\{FL\{TR\}\}, ANN\{FL\{LAG\}\}, ANN\{FL\{CH\}\}, ANN\{FL\{LE\}\}, ANN\{FL\{CE\}\}, ANN\{FL\{RCE\}\}, ANN\{RBF\}, ANN\{WAV\} & RMSE, MAPE \\ \hline
    S.-H. Hsu, Hsieh, Chih \& K.-C. Hsu, 2009 & Nikkei 225 $|$ All Ordinaries $|$ Hang Seng $|$ Straits Times $|$ TAIEX $|$ KOSPI $|$ Dow Jones (index) & Various & 1997-2002 (d) & TECH  & SOM-SVR, SVR & NMSE, MAE, DS, WDS \\ \hline
    Sagar \& Kiat, 1999 & 3 (stock) $\in$ SES & SGP   & 1996-1997 (d) & OTHER\{NEWS\} & NLP-ANN\{TD\} & MAE \\ \hline
    Sezer \& Ozbayoglu, 2018 & Dow 30 (stock) $|$ 9 ETF (ETF) & US    & 2002-2017 (d) & TECH  & ANN{C}, ANN\{R\{LSTM\}\}, ANN, BH, SMA & AR, std \\ \hline
    Shen, Tan, Zhang, Zeng \& Xu, 2018 & S\&P 500 (index) & US    & 1991-2017 (d) &  & ANN\{R\{GRU\}\}, ANN\{R\{GRU\}\}-SVM, ANN, SVM & A, AR \\ \hline
    Shynkevich, McGinnity, Coleman, Belatreche \& Li, 2017 & 50 (stock) $\in$ S\&P 500 & US    & 2002-2012 (d) & TECH  & SVM, ANN, kNN, BH & A, SR, WR \\ \hline
    Siekmann, Kruse, Gebhardt, van Overbeek \& Cooke, 2001 & DAX (index) & GER   & 1994-1998 &  & NFUZZ, LRM, NAIVE, BH & A\{HR\}, RMSE, AR, * \\ \hline
    Soto \& Melin, 2015 & MSE (index) & MEX   & 2005-2009 (d) &  & NFUZZ\{ANFIS\{Type2\}\}, NFUZZ\{ANFIS\{Type1\}\} & RMSE \\ \hline
    Steiner \& Wittkemper, 1997 & 31 (stock) $\in$ FSE & GER   & 1991-1994 & MARKET, TECH** & BH, ANN\{P\}-ANN\{GR\} & alpha, AR, std \\ \hline
    Tay \& Cao, 2001 & 5 (future) $\in$ CMM & US    & 1992-1999 (d) & TECH  & ANN, SVM & NMSE, MAE, DS, WDS \\ \hline
    Tenti, 1996 & (future) -$>$ DM & GER   & 1990-1994 (d) & TECH  & ANN\{R\} & NMSE, ROE, ROC, A \\ \hline
    Thawornwong \& Enke, 2004 & S\&P 500 (index) & US    & 1976-1999 (m) & FUND, MACRO, MARKET & ANN, ANN\{P\}, LRM, BH, RW & A, AR, std, SR, * \\ \hline
    Ticknor, 2013 & AAPL $|$ IBM (stock) & US    & 2003-2005 (d) & TECH  & ANN\{BR\}, [ANN-HMM]$<$-GA-WA, ARIMA & MAPE  \\ \hline
    Tsaih, Hsu \& Lai, 1998 & (future) -$>$ S\&P 500 & US    & 1983-1993 (d) & TECH  & ANN\{REAS\}, BH & AR, A \\ \hline
    Tseng, Cheng, Wang \& Peng, 2008 & x (option) $\in$ TXO & TW    & 2005-2006 (d) & TECH, MARKET & GARCH\{E\}-ANN, Grey-GARCH\{E\}-ANN & RMSE, MAE, MAPE \\ \hline
    Vanstone, Finnie \& Tan, 2005 & x (stock) & AUS   & 2002-2003 & FUND  & ANN, BH & AR, SR, MD, UI, * \\ \hline
    Versace, Bhatt, Hinds \& Shiffer, 2004 & DIA (ETF) & US    & 2001-2003 (d) & MARKET, MACRO, TECH & GA-$>$ANN & A \\ \hline
    Wah \& Qian, 2002 & 3 (stock) & US    & 1997-2002 (d) &  & ANN\{R\{FIR\}\}, CC, AR, ANN, IP & NMSE \\ \hline
    Walczak, 1999 & DBS50 $|$ DJIA $|$ Nikkei 225 (index) & SGP, US, JP & 1994-1995 (d) & MARKET, TECH & ANN   & A \\ \hline
    Wang \& Chan, 2006 & 3 (stock) & US    & 1990-1996 (d) & TECH  & DT\{BIAS\} & AR, std, A \\ \hline
    Wang, 2009 & x (option) $\in$ TXO & TW    & 2005-2006 (d) &  & Grey-GARCH\{GJR\}-ANN, GARCH\{GJR\}-ANN, GARCH-ANN & RMSE, MAE, MAPE \\ \hline
    Wang, Xu \& Zheng, 2018 & SSE (index) & CN    & 2012-2015 (d) & OTHER\{NEWS\}, TECH & RSE-DBN, ANN, SVM, RF, ANN\{R\}, ANN\{R\{LSTM\}\} & F1, Precision, Recall Accuracy, AUC \\ \hline
    Wen, Yang \& Song, 2009 & S\&P 500 (index) & US    & 1000d &  & NFUZZ\{ANFIS\}, ANN, SVM, NFUZZ\{ANFIS\}\^{}ANN\^{}SVM, [NFUZZ\{ANFIS\}\^{}ANN\^{}SVM]-ANN & MSE \\ \hline
    Wen, Yang, Song \& Jia, 2010 & 422 (stock) $\in$ S\&P 500  $|$ MSFT $|$ IBM (stock) & US    & 11-12y** $|$ 2004-2008 (d)\{OT\} $|$ 2004-2008 (d)\{OT\} & TECH  & SVM, BH & AR, SCC, MSE \\ \hline
    Witkowska, 1995 & 3 (stock) $\in$ PSE & PL    & 1993 & MARKET & ANN   & MSE \\ \hline
    Wittkemper \& Steiner, 1996 & 67 (stock) & GER   & 1967-1986 (d) & FUND, TECH & ANN\{GR\}, ANN\{GR\}$<$-GA & MSE \\ \hline
    Wu, Fung \& Flitman, 2001 & S\&P 500 (index) & US    & 1992-2000 (m) & MACRO & NFUZZ\{FFNF\}, ANN & A \\ \hline
    Yeh, Lien \& Tsai, 2011 & TAIEX (index) & TW    & 1989-2004 (d) & TECH  & ANN   & AR \\ \hline
    Yu, Wang \& Lai, 2008 & S\&P 500 $|$ FTSE100 $|$ Nikkei 225 (index) $|$ USD/EUR $|$ USD/GBP $|$ USD/JPY (FX) & Various & 2000-2003 (d) &  & RW, AES, ARIMA, ANN & alpha \\ \hline
    Yumlu, Gurgen \& Okay, 2004 & XU100 (index) & TR    & 1989-2003 (d) &  & ENS[ANN], ANN\{R\}, GARCH\{GJR\} & ThIC, CC, A\{HR\}, MSE \\ \hline
    Yumlu, Gurgen \& Okay, 2005 & XU100 (index) & TR    & 1990-2002 (d) &  & ANN, ANN\{R\{EL\}\}, MOE, GARCH\{E\} & A\{HR\}, A\{HR+\}, A\{HR-\}, MSE, MAE, ρ \\ \hline
    Zhang \& Berardi, 2001 & GBP/USD (FX) & UK, US & 1976-1994 (w) &  & ENS\{SYS\}[ANN], ENS\{SER\}[ANN] & MSE, MAE \\ \hline
    Zhang \& Pan, 2016 & 20 (stock) $\in$ SZSE $|$ 16 (stock) $\in$ NASDAQ & CN, US & 2010-2015 & TECH  & SVM\{P\}$<$-AdaBoost$<$-GA, SVM\{P\}, ANN & A, g-means \\ \hline
    Zhang \& Wan, 2007 & USD/JPY $|$ USD/GBP $|$ USD/HKD (FX) & Various & 1998-2001 &  & NFUZZ\{SFI\} & MAPE \\ \hline
    Zhang, 2003 & GBP/USD (FX) & UK, US & 1980-1993 (w) &  & ANN\^{}ARIMA, ANN, ARIMA & MSE, MAE \\ \hline
    Zhang, Akkaladevi, Vachtsevanos \& Lin, 2002 & 6 (stock)  & US    & 1981-1994 &  & NFUZZ\{GNN\}, ANN & MAE** \\ \hline
    Zhang, Jiang \& Li, 2004 & SCI (index) & CN    & 1995-2003 (d) &  & ANN, BH & R, A \\ \hline
    Zhu, Wang, Xu \& Li, 2008 & NASDAQ $|$ DJIA $|$ STI (index) & US, SGP & 1997-2005 $|$ 1990-2005 $|$ 1989-2005 & MARKET, TECH & ANN   & A, MSE \\ \hline
\end{tabularx}
\endgroup
\clearpage
\twocolumn
\captionsetup{font={sc}}
\tablecaption*{Algorithm Abbreviations}
\begin{small}
\twocolumn
\begin{supertabular}{ll}
A     & Adaptive parameters \\
AA    & Auto-associative \\
ABC   & Artificial Bee Colony Algorithm \\
AE    & Autoencoder \\
AES   & Adaptive Exponential Smoothing \\
ALC   & Adaptive Linear Combiner \\
ANFIS & Adaptive Network-based Fuzzy Inference System \\
ANN   & Artificial Neural Network \\
AP    & Asymmetric Power \\
AR    & Autoregressive \\
ARFIMA & Autoregressive Fractionally Integrated Moving Average \\
ARIMA & Autoregressive Integrated Moving Average \\
ARMA  & Autoregressive Moving Average \\
ARMR  & Autoregressive Moving Reference \\
ATD   & Adaptive Time Delay \\
B     & Basic \\
BART  & Bayesian Autoregressive Tree \\
BFO   & Bacterial Foraging Optimization \\
BH    & Buy-and-hold \\
bi    & Bivariate \\
BM    & Boltzmann Machine \\
BR    & Bayesian Regularized \\
BVAR  & Bayesian Vector Autoregression \\
C     & Convolutional \\
CBE   & Context-Based Ensemble \\
CBR   & Case-Based Reasoning \\
CC    & Carbon Copy \\
CCA   & Curvilinear Component Analysis \\
CE    & Computationally Efficient \\
CFL   & Cascaded Functional Link \\
CFS   & Correlation-based Feature Selection \\
CH    & Chebyshev polynomials \\
CLS   & Constrained Least Squares \\
CMAC  & Cerebellar Model Articulation Controller \\
CN    & Control Nodes \\
CS    & Class-Sensitive \\
D     & Deep \\
DA    & Dynamic Adaptive \\
DAN2  & Dynamic Architecture for artificial neural Networks \\
DB    & Distance-Based \\
DBN   & Deep Belief Network \\
DC    & Dynamic Cognitive \\
DENFIS & Dynamic Evolving Neural-Fuzzy Inference System \\
DOE   & Design Of Experiment \\
DRP   & Dynamic Ridge Polynomial \\
DT    & Decision Tree \\
DWT   & Discrete Wavelet Transform \\
E     & Exponential \\
EFNN  & Evolving Fuzzy Neural Network \\
EL    & Elman \\
ELM   & Extreme Learning Machine \\
ENS   & Ensemble \\
ES    & Exponential Smoothing \\
FFNF  & Feed Forward Neuro Fuzzy \\
FI    & Fractionally Integrated \\
FIAP  & Fractionally Integrated Asymmetric Power \\
FIR   & Finite-duration Impulse Response \\
FIS   & Fuzzy Inference System \\
FL    & Functional Link \\
FNT   & Flexible Neural Tree \\
FSSFS & F-score and Supported Sequential Forward Search  \\
FTS   & Fuzzy Time Series \\
FUZZ  & Fuzzy Logic \\
G     & Generalized \\
GA    & Genetic Algorithm \\
GARCH & General Autoregressive Conditional Heteroskedasticity \\
GB    & Gradient-Boosted \\
GGAP  & General Growing And Pruning \\
GJR   & Glosten-Jagannathan-Runkle \\
GM    & Gaussian Mixture \\
GMM   & Generalized Method of Moments \\
GNN   & Granular Neural Network \\
GNP   & Genetic Network Programming \\
GR    & General Regression \\
GRU   & Gated Recurrent Unit \\
H     & Hopfield \\
HAR   & Heterogeneous Autoregressive \\
HE    & Hybrid Elman \\
HMM   & Hidden Markov Model \\
HO    & Higher Order \\
HiCEFS & Hierarchical Co-Evolutionary Fuzzy System \\
IDL   & Increasing-Decreasing-Linear \\
IG    & Information Gain \\
IP    & Ideal Predictor \\
J     & Jumps \\
KF    & Kalman Filter \\
L     & Linear \\
LAG   & Laguerre polynomials \\
LC    & Linear Correlation \\
LDA   & Linear Discriminant Analysis \\
LE    & Legendre polynomials \\
LICA  & Linear Independent Component Analysis \\
LLWAV & Local Linear Wavelet \\
LMS   & Least Mean Squares \\
LOGIT & Logistic Regression \\
LRM   & Linear Regression Model \\
LST   & Logistic Smooth Transition \\
LSTAR & Logistic Smooth Transition Autoregressive \\
LSTM  & Long Short-Term Memory \\
LTS   & Lagged Time Series \\
LWPR  & Local Weighted Polynomial Regression \\
M     & Mamdani \\
MACDM & Moving Average Convergence Divergence Model \\
MAJ   & Majority \\
MOE   & Mixture of Experts \\
MSNR  & Microstructure Noise Robust \\
MTF   & Multivariate Transfer Function \\
MarkS & Markov Switching \\
N     & Nonlinear \\
NBAY  & Naive Bayes \\
NFUZZ & Neuro-Fuzzy \\
NLICA & Nonlinear Independent Component Analysis \\
NLM   & Non-Linear Model \\
NLP   & Natural Language Processing \\
NLRM  & Non-Linear Regression Model \\
NP    & Nonlinear Power \\
NXCS  & Neural Extended Classifier System \\
OLS   & Ordinary Least Squares \\
P     & Probabilistic \\
PC    & Partially Connected \\
PCA   & Principal Components Analysis \\
PLR   & Piecewise Linear Representation \\
POLY  & Polynomial \\
POW   & Power \\
PP    & Polynomial Pipelined \\
PROX  & Proximal \\
PS    & Psi Sigma \\
PSO   & Particle Swarm Optimization \\
PWNL  & Piecewise Nonlinear \\
QDA   & Quadratic Discriminant Analysis \\
QP    & Quasiperiodic \\
R     & Recurrent \\
RA    & Resource Allocation \\
RBF   & Radial Basis Function \\
RBM   & Restricted Boltzmann Machine \\
RC    & Rank Correlation \\
RCE   & Recurrent Computationally Efficient \\
REAS  & Reasoning \\
RF    & Random Forest \\
RL    & Reinforcement Learning \\
RLSF  & Regularized Least Squares Fuzzy \\
RMD   & Recurrent Mixture Density \\
RP    & Ridge Polynomial \\
RR    & Regression Relief \\
RSE   & Random Subspace Ensemble \\
RSPOP & Rough Set-based Pseudo Outer-Product \\
RW    & Random Walk \\
S     & Static Ensemble \\
SA    & Simple Asymmetric \\
SCE   & Softmax Cross Entropy \\
SER   & Serial \\
SFI   & Statistical Fuzzy Interval \\
SH    & Sell and Hold \\
SMA   & Simple Moving Average \\
SO    & Self-Organizing \\
SOM   & Self-Organizing Map \\
STAR  & Smooth Transition Autoregressive \\
STE   & Stochastic Time Effective \\
SU    & Symmetrical Uncertainty \\
SVM   & Support Vector Machine \\
SVR   & Support Vector Regression \\
SYS   & Systematic \\
SimA  & Simulated Annealing \\
T     & Threshold \\
TAR   & Threshold Autoregressive \\
TD    & Time Delay \\
TR    & Trigonometric Funcion \\
TS    & Tabu Search \\
TSK   & Takagi-Sugeno-Kang \\
Type1 & Type 1 Fuzzy Logic \\
Type2 & Type 2 Fuzzy Logic \\
V     & Voting \\
VAR   & Variance-based \\
VEC   & Vector Error Correction \\
W     & Weighted \\
WA    & Weighted Average \\
WAV   & Wavelet \\
WBP   & Weighted Backpropagation \\
kNN   & k-Nearest Neighbors \\
multi & Multivariate \\
subs  & Substitutes \\
uni   & Univariate \\

\end{supertabular}
\end{small}


\begin{thebibliography}{00}
\bibitem{} Abraham, A., Nath, B., \& Mahanti, P. K. (2001, May). Hybrid intelligent systems for stock market analysis. In International Conference on Computational Science (pp. 337-345). Springer, Berlin, Heidelberg.
\bibitem{} Adhikari, R., \& Agrawal, R. K. (2014). A combination of artificial neural network and random walk models for financial time series forecasting. Neural Computing and Applications, 24(6), 1441-1449.
\bibitem{} Andreou, A. S., Neocleous, C. C., Schizas, C. N., \& Toumpouris, C. (2000). Testing the predictability of the Cyprus Stock Exchange: The case of an emerging market. In Neural Networks, 2000. IJCNN 2000, Proceedings of the IEEE-INNS-ENNS International Joint Conference on (Vol. 6, pp. 360-365). IEEE.
\bibitem{} Araujo, R. D. A., Nedjah, N., de Seixas, J. M., Oliveira, A. L., \& Silvio, R. D. L. (2018). Evolutionary-morphological learning machines for high-frequency financial time series prediction. Swarm and Evolutionary Computation, 42, 1-15.
\bibitem{} Armano, G., Marchesi, M., \& Murru, A. (2005). A hybrid genetic-neural architecture for stock indexes forecasting. Information Sciences, 170(1), 3-33.
\bibitem{} Atiya, A., Talaat, N., \& Shaheen, S. (1997, June). An efficient stock market forecasting model using neural networks. In Neural Networks, 1997., International Conference on (Vol. 4, pp. 2112-2115). IEEE.
\bibitem{atsalakis} Atsalakis, G. S., \& Valvanis, K. P. (2009). Surveying stock market forecasting techniques - Part II: Soft computing methods. Expert Systems with Applications, 36, 5932-5941. 
\bibitem{} Bagheri, A., Peyhani, H. M., \& Akbari, M. (2014). Financial forecasting using ANFIS networks with quantum-behaved particle swarm optimization. Expert Systems with Applications, 41(14), 6235-6250.
\bibitem{bahrammirzaee} Bahrammirzaee, A. (2010). A comparative survey of artificial intelligence applications in finance: artificial neural networks, expert system and hybrid intelligent systems. Neural Computing and Applications, 19(8), 1165-1195.
\bibitem{} Banik, S., Chanchary, F. H., Rouf, R. A., \& Khan, K. (2007, December). Modeling chaotic behavior of Dhaka Stock Market Index values using the neuro-fuzzy model. In Computer and information technology, 2007. iccit 2007. 10th international conference on (pp. 1-6). IEEE.
\bibitem{} Bildirici, M., \& Ersin, O. O. (2009). Improving forecasts of GARCH family models with the artificial neural networks: An application to the daily returns in Istanbul Stock Exchange. Expert Systems with Applications, 36(4), 7355-7362.
\bibitem{} Bildirici, M., \& Ersin, O. O. (2013). Forecasting oil prices: Smooth transition and neural network augmented GARCH family models. Journal of Petroleum Science and Engineering, 109, 230-240.
\bibitem{} Bildirici, M., Alp, E. A., \& Ersin, O. O. (2010). TAR-cointegration neural network model: An empirical analysis of exchange rates and stock returns. Expert Systems with Applications, 37(1), 2-11.
\bibitem{} Bodyanskiy, Y., \& Popov, S. (2006). Neural network approach to forecasting of quasiperiodic financial time series. European Journal of Operational Research, 175(3), 1357-1366.
\bibitem{} Cao, L. J., \& Tay, F. E. H. (2003). Support vector machine with adaptive parameters in financial time series forecasting. IEEE Transactions on neural networks, 14(6), 1506-1518.
\bibitem{} Cao, L., \& Tay, F. E. (2001). Financial forecasting using support vector machines. Neural Computing \& Applications, 10(2), 184-192.
\bibitem{} Cao, Q., Leggio, K. B., \& Schniederjans, M. J. (2005). A comparison between Fama and French's model and artificial neural networks in predicting the Chinese stock market. Computers \& Operations Research, 32(10), 2499-2512.
\bibitem{} Casas, C. A. (2001). Tactical asset allocation: an artificial neural network based model. In Neural Networks, 2001. Proceedings. IJCNN'01. International Joint Conference on (Vol. 3, pp. 1811-1816). IEEE.
\bibitem{cavalcante} Cavalcante, R. C., Brasileiro, R. C., Souza, V. L., Nobrega, J. P., \& Oliveira, A. L. (2016). Computational intelligence and financial markets: A survey and future directions. Expert Systems with Applications, 55, 194-211.
\bibitem{} Chang, P. C. (2012). A novel model by evolving partially connected neural network for stock price trend forecasting. Expert Systems with Applications, 39(1), 611-620.
\bibitem{} Chang, P. C., \& Liu, C. H. (2008). A TSK type fuzzy rule based system for stock price prediction. Expert Systems with Applications, 34(1), 135-144.
\bibitem{} Chang, P. C., Fan, C. Y., \& Liu, C. H. (2009). Integrating a piecewise linear representation method and a neural network model for stock trading points prediction. IEEE Transactions on Systems, Man, and Cybernetics, Part C (Applications and Reviews), 39(1), 80-92.
\bibitem{} Chang, P. C., Liu, C. H., Lin, J. L., Fan, C. Y., \& Ng, C. S. (2009). A neural network with a case based dynamic window for stock trading prediction. Expert Systems with Applications, 36(3), 6889-6898.
\bibitem{} Chavarnakul, T., \& Enke, D. (2008). Intelligent technical analysis based equivolume charting for stock trading using neural networks. Expert Systems with Applications, 34(2), 1004-1017.
\bibitem{} Chen, A. S., \& Leung, M. T. (2004). Regression neural network for error correction in foreign exchange forecasting and trading. Computers \& Operations Research, 31(7), 1049-1068.
\bibitem{} Chen, A. S., Leung, M. T., \& Daouk, H. (2003). Application of neural networks to an emerging financial market: forecasting and trading the Taiwan Stock Index. Computers \& Operations Research, 30(6), 901-923.
\bibitem{} Chen, C. H. (1994, June). Neural networks for financial market prediction. In Neural Networks, 1994. IEEE World Congress on Computational Intelligence., 1994 IEEE International Conference on (Vol. 2, pp. 1199-1202). IEEE.
\bibitem{} Chen, M. Y., Chen, D. R., Fan, M. H., \& Huang, T. Y. (2013). International transmission of stock market movements: an adaptive neuro-fuzzy inference system for analysis of TAIEX forecasting. Neural Computing and Applications, 23(1), 369-378.
\bibitem{} Chen, M. Y., Fan, M. H., Chen, Y. L., \& Wei, H. M. (2013). Design of experiments on neural network's parameters optimization for time series forecasting in stock markets. Neural Network World, 23(4), 369.
\bibitem{} Chen, W. H., Shih, J. Y., \& Wu, S. (2006). Comparison of support-vector machines and back propagation neural networks in forecasting the six major Asian stock markets. International Journal of Electronic Finance, 1(1), 49-67.
\bibitem{} Chen, Y., Abraham, A., Yang, J., \& Yang, B. (2005, August). Hybrid methods for stock index modeling. In International Conference on Fuzzy Systems and Knowledge Discovery (pp. 1067-1070). Springer, Berlin, Heidelberg.
\bibitem{} Chen, Y., Dong, X., \& Zhao, Y. (2005, October). Stock index modeling using EDA based local linear wavelet neural network. In Neural Networks and Brain, 2005. ICNN\&B'05. International Conference on (Vol. 3, pp. 1646-1650). IEEE.
\bibitem{} Chen, Y., Ohkawa, E., Mabu, S., Shimada, K., \& Hirasawa, K. (2009). A portfolio optimization model using Genetic Network Programming with control nodes. Expert Systems with Applications, 36(7), 10735-10745.
\bibitem{} Chen, Y., Yang, B., \& Abraham, A. (2007). Flexible neural trees ensemble for stock index modeling. Neurocomputing, 70(4-6), 697-703.
\bibitem{} Chenoweth, T., \& Obradović, Z. (1996). A multi-component nonlinear prediction system for the S\&P 500 index. Neurocomputing, 10(3), 275-290.
\bibitem{} Chenoweth, T., Obradovic, Z., \& Lee, S. S. (1996). Embedding technical analysis into neural network based trading systems. Applied Artificial Intelligence, 10(6), 523-542.
\bibitem{} Chiang, W. C., Urban, T. L., \& Baldridge, G. W. (1996). A neural network approach to mutual fund net asset value forecasting. Omega, 24(2), 205-215.
\bibitem{} Chong, E., Han, C., \& Park, F. C. (2017). Deep learning networks for stock market analysis and prediction: Methodology, data representations, and case studies. Expert Systems with Applications, 83, 187-205.
\bibitem{} Chun, S. H., \& Park, Y. J. (2005). Dynamic adaptive ensemble case-based reasoning: application to stock market prediction. Expert Systems with Applications, 28(3), 435-443.
\bibitem{GRU} Chung, J., Gulcehre, C., Cho, K., \& Bengio, Y. (2014). Empirical evaluation of gated recurrent neural networks on sequence modeling. arXiv preprint arXiv:1412.3555.
\bibitem{} Constantinou, E., Georgiades, R., Kazandjian, A., \& Kouretas, G. P. (2006). Regime switching and artificial neural network forecasting of the Cyprus Stock Exchange daily returns. International Journal of Finance \& Economics, 11(4), 371-383.
\bibitem{} Dai, W., Wu, J. Y., \& Lu, C. J. (2012). Combining nonlinear independent component analysis and neural network for the prediction of Asian stock market indexes. Expert Systems with applications, 39(4), 4444-4452.
\bibitem{Daniskova} Daniskova, K., \& Fidrmuc, J. (2012). Meta-Analysis of the New Keynesian Phillips Curve (No. 314). IOS Working Papers.
\bibitem{} Das, S. R., Mishra, D., \& Rout, M. (2017). A hybridized ELM-Jaya forecasting model for currency exchange prediction. Journal of King Saud University-Computer and Information Sciences.
\bibitem{} Dash, R., \& Dash, P. K. (2016). A hybrid stock trading framework integrating technical analysis with machine learning techniques. The Journal of Finance and Data Science, 2(1), 42-57.
\bibitem{} De Faria, E. L., Albuquerque, M. P., Gonzalez, J. L., Cavalcante, J. T. P., \& Albuquerque, M. P. (2009). Predicting the Brazilian stock market through neural networks and adaptive exponential smoothing methods. Expert Systems with Applications, 36(10), 12506-12509.
\bibitem{} Dempster, M. A., \& Leemans, V. (2006). An automated FX trading system using adaptive reinforcement learning. Expert Systems with Applications, 30(3), 543-552.
\bibitem{} Doeksen, B., Abraham, A., Thomas, J., \& Paprzycki, M. (2005, April). Real stock trading using soft computing models. In International Conference on Information Technology: Coding and Computing (ITCC'05)-Volume II (Vol. 2, pp. 162-167). IEEE.
\bibitem{} Dunis, C. L., Laws, J., \& Sermpinis, G. (2011). Higher order and recurrent neural architectures for trading the EUR/USD exchange rate. Quantitative Finance, 11(4), 615-629.
\bibitem{Elman} Elman, J. L. (1990). Finding structure in time. Cognitive science, 14(2), 179-211.
\bibitem{} Enke, D., \& Thawornwong, S. (2005). The use of data mining and neural networks for forecasting stock market returns. Expert Systems with applications, 29(4), 927-940.
\bibitem{} Fatima, S., \& Hussain, G. (2008). Statistical models of KSE100 index using hybrid financial systems. Neurocomputing, 71(13-15), 2742-2746.
\bibitem{Fernandez-Rodriguez} Fernandez-Rodrıguez, F., Gonzalez-Martel, C., \& Sosvilla-Rivero, S. (2000). On the profitability of technical trading rules based on artificial neural networks: Evidence from the Madrid stock market. Economics letters, 69(1), 89-94.
\bibitem{} Fernandez, A., \& Gomez, S. (2007). Portfolio selection using neural networks. Computers \& Operations Research, 34(4), 1177-1191.
\bibitem{} Fischer, T., \& Krauss, C. (2018). Deep learning with long short-term memory networks for financial market predictions. European Journal of Operational Research, 270(2), 654-669.
\bibitem{} Freitas, F. D., De Souza, A. F., \& de Almeida, A. R. (2009). Prediction-based portfolio optimization model using neural networks. Neurocomputing, 72(10-12), 2155-2170.
\bibitem{} Ghazali, R., Hussain, A. J., Nawi, N. M., \& Mohamad, B. (2009). Non-stationary and stationary prediction of financial time series using dynamic ridge polynomial neural network. Neurocomputing, 72(10-12), 2359-2367.
\bibitem{Glass} Glass, G. V. (1976). Primary, secondary, and meta-analysis of research. Educational researcher, 5(10), 3-8.
\bibitem{} Grudnitski, G., \& Osburn, L. (1993). Forecasting S\&P and gold futures prices: An application of neural networks. Journal of Futures Markets, 13(6), 631-643.
\bibitem{} Guresen, E., Kayakutlu, G., \& Daim, T. U. (2011). Using artificial neural network models in stock market index prediction. Expert Systems with Applications, 38(8), 10389-10397.
\bibitem{} Guresen, E., \& Kayakutlu, G. (2008, October). Forecasting stock exchange movements using artificial neural network models and hybrid models. In International Conference on Intelligent Information Processing (pp. 129-137). Springer, Boston, MA.
\bibitem{} Hajizadeh, E., Seifi, A., Zarandi, M. F., \& Turksen, I. B. (2012). A hybrid modeling approach for forecasting the volatility of S\&P 500 index return. Expert Systems with Applications, 39(1), 431-436.
\bibitem{} Harvey, C. R., Travers, K. E., \& Costa, M. J. (2000). Forecasting emerging market returns using neural networks. Emerging Markets Quarterly, 4, 43-54.
\bibitem{} Hassan, M. R. (2009). A combination of hidden Markov model and fuzzy model for stock market forecasting. Neurocomputing, 72(16-18), 3439-3446.
\bibitem{} Hassan, M. R., Nath, B., \& Kirley, M. (2007). A fusion model of HMM, ANN and GA for stock market forecasting. Expert Systems with Applications, 33(1), 171-180.
\bibitem{Haykin} Haykin, S. (1994). Neural networks: a comprehensive foundation. Prentice Hall PTR.
\bibitem{LSTM} Hochreiter, S., \& Schmidhuber, J. (1997). Long short-term memory. Neural computation, 9(8), 1735-1780.
\bibitem{} Hsieh, T. J., Hsiao, H. F., \& Yeh, W. C. (2011). Forecasting stock markets using wavelet transforms and recurrent neural networks: An integrated system based on artificial bee colony algorithm. Applied soft computing, 11(2), 2510-2525.
\bibitem{} Hsu, S. H., Hsieh, J. P. A., Chih, T. C., \& Hsu, K. C. (2009). A two-stage architecture for stock price forecasting by integrating self-organizing map and support vector regression. Expert Systems with Applications, 36(4), 7947-7951.
\bibitem{Huang1} Huang, C. L., \& Tsai, C. Y. (2009). A hybrid SOM-SVR with a filter-based feature selection for stock market forecasting. Expert Systems with Applications, 36(2), 1529-1539.
\bibitem{Huang2} Huang, H., Pasquier, M., \& Quek, C. (2009). Financial market trading system with a hierarchical coevolutionary fuzzy predictive model. IEEE Transactions on Evolutionary Computation, 13(1), 56-70.
\bibitem{} Huang, S. C., \& Wu, T. K. (2008). Integrating GA-based time-scale feature extractions with SVMs for stock index forecasting. Expert Systems with Applications, 35(4), 2080-2088.
\bibitem{} Huang, W., Nakamori, Y., \& Wang, S. Y. (2005). Forecasting stock market movement direction with support vector machine. Computers \& Operations Research, 32(10), 2513-2522.
\bibitem{} Hussain, A. J., Knowles, A., Lisboa, P. J., \& El-Deredy, W. (2008). Financial time series prediction using polynomial pipelined neural networks. Expert Systems with Applications, 35(3), 1186-1199.
\bibitem{Jarrell} Jarrell, S. B., \& Stanley, T. D. (1990). A meta-analysis of the union-nonunion wage gap. ILR Review, 44(1), 54-67.
\bibitem{} Kanas, A., \& Yannopoulos, A. (2001). Comparing linear and nonlinear forecasts for stock returns. International Review of Economics \& Finance, 10(4), 383-398.
\bibitem{} Kara, Y., Boyacioglu, M. A., \& Baykan, O. K. (2011). Predicting direction of stock price index movement using artificial neural networks and support vector machines: The sample of the Istanbul Stock Exchange. Expert Systems with Applications, 38(5), 5311-5319.
\bibitem{} Khan, M. A. I. (2011). Financial volatility forecasting by nonlinear support vector machine heterogeneous autoregressive model: evidence from Nikkei 225 stock index. International Journal of Economics and Finance, 3(4), 138.
\bibitem{khadjeh} Khadjeh Nassirtoussi, A., Aghabozorgi, S., Wah, T. Y., Ngo, D. C. L. (2015). Text mining of news headlines for FOREX market prediction: A Multi-layer Dimension Reduction Algorithm with semantics and sentiment. Expert Systems with Applications, 42, 306-324. 
\bibitem{} Khemchandani, R., \& Chandra, S. (2009). Regularized least squares fuzzy support vector regression for financial time series forecasting. Expert Systems with Applications, 36(1), 132-138.
\bibitem{} Kim, H. J., \& Shin, K. S. (2007). A hybrid approach based on neural networks and genetic algorithms for detecting temporal patterns in stock markets. Applied Soft Computing, 7(2), 569-576.
\bibitem{} Kim, K. J. (2003). Financial time series forecasting using support vector machines. Neurocomputing, 55(1-2), 307-319.
\bibitem{} Kim, K. J. (2006). Artificial neural networks with evolutionary instance selection for financial forecasting. Expert Systems with Applications, 30(3), 519-526.
\bibitem{} Kim, K. J., \& Ahn, H. (2012). Simultaneous optimization of artificial neural networks for financial forecasting. Applied Intelligence, 36(4), 887-898.
\bibitem{} Kim, K. J., \& Han, I. (2000). Genetic algorithms approach to feature discretization in artificial neural networks for the prediction of stock price index. Expert Systems with Applications, 19(2), 125-132.
\bibitem{} Kim, K. J., Han, I., \& Chandler, J. S. (1998). Extracting trading rules from the multiple classifiers and technical indicators in stock market. In The Korea Society of Management Information Systems' 98 International Conference on IS Paradigm reestablishment. The Korea Society of Management Information Systems.
\bibitem{} Kim, S. H., \& Chun, S. H. (1998). Graded forecasting using an array of bipolar predictions: application of probabilistic neural networks to a stock market index. International Journal of Forecasting, 14(3), 323-337.
\bibitem{} Kimoto, T., Asakawa, K., Yoda, M., \& Takeoka, M. (1990, June). Stock market prediction system with modular neural networks. In Neural Networks, 1990., 1990 IJCNN International Joint Conference on (pp. 1-6). IEEE.
\bibitem{} Ko, P. C., \& Lin, P. C. (2008). Resource allocation neural network in portfolio selection. Expert Systems with Applications, 35(1-2), 330-337.
\bibitem{} Koulouriotis, D. E., Diakoulakis, I. E., Emiris, D. M., \& Zopounidis, C. D. (2005). Development of dynamic cognitive networks as complex systems approximators: validation in financial time series. Applied Soft Computing, 5(2), 157-179.
\bibitem{} Krauss, C., Do, X. A., \& Huck, N. (2017). Deep neural networks, gradient-boosted trees, random forests: Statistical arbitrage on the S\&P 500. European Journal of Operational Research, 259(2), 689-702.
\bibitem{} Kristjanpoller, W., \& Minutolo, M. C. (2015). Gold price volatility: A forecasting approach using the Artificial Neural Network-GARCH model. Expert Systems with Applications, 42(20), 7245-7251.
\bibitem{} Kristjanpoller, W., \& Minutolo, M. C. (2016). Forecasting volatility of oil price using an artificial neural network-GARCH model. Expert Systems with Applications, 65, 233-241.
\bibitem{} Kristjanpoller, W., \& Minutolo, M. C. (2018). A hybrid volatility forecasting framework integrating GARCH, artificial neural network, technical analysis and principal components analysis. Expert Systems with Applications, 109, 1-11.
\bibitem{} Kristjanpoller, W., Fadic, A., \& Minutolo, M. C. (2014). Volatility forecast using hybrid neural network models. Expert Systems with Applications, 41(5), 2437-2442.
\bibitem{} Kryzanowski, L., Galler, M., \& Wright, D. W. (1993). Using artificial neural networks to pick stocks. Financial Analysts Journal, 21-27.
\bibitem{} Kumar, D., Meghwani, S. S., \& Thakur, M. (2016). Proximal support vector machine based hybrid prediction models for trend forecasting in financial markets. Journal of Computational Science, 17, 1-13.
\bibitem{} Kuo, R. J. (1998). A decision support system for the stock market through integration of fuzzy neural networks and fuzzy Delphi. Applied Artificial Intelligence, 12(6), 501-520.
\bibitem{} Kuo, R. J., Lee, L. C., \& Lee, C. F. (1996, October). Integration of artificial neural networks and fuzzy delphi for stock market forecasting. In Systems, Man, and Cybernetics, 1996., IEEE International Conference on(Vol. 2, pp. 1073-1078). IEEE.
\bibitem{} Kwon, Y. K., \& Moon, B. R. (2007). A hybrid neurogenetic approach for stock forecasting. IEEE transactions on neural networks, 18(3), 851-864.
\bibitem{} Lam, M. (2004). Neural network techniques for financial performance prediction: integrating fundamental and technical analysis. Decision support systems, 37(4), 567-581.
\bibitem{} Lee, J., Cho, S., \& Baek, J. (2003, March). Trend detection using auto-associative neural networks: Intraday KOSPI 200 futures. In Computational Intelligence for Financial Engineering, 2003. Proceedings. 2003 IEEE International Conference on (pp. 417-420). IEEE.
\bibitem{} Lee, M. C. (2009). Using support vector machine with a hybrid feature selection method to the stock trend prediction. Expert Systems with Applications, 36(8), 10896-10904.
\bibitem{} Lee, T. S., \& Chen, N. J. (2002). Investigating the information content of non-cash-trading index futures using neural networks. Expert Systems with Applications, 22(3), 225-234.
\bibitem{} Lee, T. S., \& Chiu, C. C. (2002). Neural network forecasting of an opening cash price index. International Journal of Systems Science, 33(3), 229-237.
\bibitem{} Leigh, W., Paz, M., \& Purvis, R. (2002). An analysis of a hybrid neural network and pattern recognition technique for predicting short-term increases in the NYSE composite index. Omega, 30(2), 69-76.
\bibitem{} Lendasse, A., de Bodt, E., Wertz, V., \& Verleysen, M. (2000). Non-linear financial time series forecasting-Application to the Bel 20 stock market index. European Journal of Economic and Social Systems, 14(1), 81-91.
\bibitem{} Leu, Y., Lee, C. P., \& Jou, Y. Z. (2009). A distance-based fuzzy time series model for exchange rates forecasting. Expert Systems with Applications, 36(4), 8107-8114.
\bibitem{} Li, S. T., \& Kuo, S. C. (2008). Knowledge discovery in financial investment for forecasting and trading strategy through wavelet-based SOM networks. Expert Systems with applications, 34(2), 935-951.
\bibitem{} Li, X., Zhang, Y., Wong, H. S., \& Qin, Z. (2009). A hybrid intelligent algorithm for portfolio selection problem with fuzzy returns. Journal of Computational and Applied Mathematics, 233(2), 264-278.
\bibitem{} Liao, Z., \& Wang, J. (2010). Forecasting model of global stock index by stochastic time effective neural network. Expert Systems with Applications, 37(1), 834-841.
\bibitem{} Lin, C. T., \& Yeh, H. Y. (2009). Empirical of the Taiwan stock index option price forecasting model-applied artificial neural network. Applied Economics, 41(15), 1965-1972.
\bibitem{} Lu, C. J., \& Wu, J. Y. (2011). An efficient CMAC neural network for stock index forecasting. Expert Systems with Applications, 38(12), 15194-15201.
\bibitem{} Lu, X., Que, D., \& Cao, G. (2016). Volatility forecast based on the hybrid artificial neural network and GARCH-type models. Procedia Computer Science, 91, 1044-1049.
\bibitem{} Majhi, R., Panda, G., \& Sahoo, G. (2009). Efficient prediction of exchange rates with low complexity artificial neural network models. Expert Systems with applications, 36(1), 181-189.
\bibitem{} Majhi, R., Panda, G., Sahoo, G., Dash, P. K., \& Das, D. P. (2007, September). Stock market prediction of S\&P 500 and DJIA using bacterial foraging optimization technique. In Evolutionary Computation, 2007. CEC 2007. IEEE Congress on (pp. 2569-2575). IEEE.
\bibitem{} Majhi, R., Panda, G., Sahoo, G., Panda, A., \& Choubey, A. (2008, June). Prediction of S\&P 500 and DJIA stock indices using particle swarm optimization technique. In 2008 IEEE Congress on Evolutionary Computation (IEEE World Congress on Computational Intelligence) (pp. 1276-1282). IEEE.
\bibitem{} Malliaris, M., \& Salchenberger, L. (1993, March). Beating the best: A neural network challenges the Black-Scholes formula. In Artificial Intelligence for Applications, 1993. Proceedings., Ninth Conference on (pp. 445-449). IEEE.
\bibitem{Mendel} Mendel, J. M., \& McLaren, R. W. (1970). 8 Reinforcement-Learning Control and Pattern Recognition Systems. In Mathematics in Science and Engineering (Vol. 66, pp. 287-318). Elsevier.
\bibitem{} Mizuno, H., Kosaka, M., Yajima, H., \& Komoda, N. (1998). Application of neural network to technical analysis of stock market prediction. Studies in Informatic and control, 7(3), 111-120.
\bibitem{} Monfared, S. A., \& Enke, D. (2014). Volatility forecasting using a hybrid GJR-GARCH neural network model. Procedia Computer Science, 36, 246-253.
\bibitem{direct_rl} Moody, J., \& Saffell, M. (2001). Learning to trade via direct reinforcement. IEEE transactions on neural Networks, 12(4), 875-889.
\bibitem{} Motiwalla, L., \& Wahab, M. (2000). Predictable variation and profitable trading of US equities: a trading simulation using neural networks. Computers \& Operations Research, 27(11-12), 1111-1129.
\bibitem{} Nayak, S. C., Misra, B. B., \& Behera, H. S. (2012, February). Index prediction with neuro-genetic hybrid network: A comparative analysis of performance. In Computing, Communication and Applications (ICCCA), 2012 International Conference on (pp. 1-6). IEEE.
\bibitem{} Ni, H., \& Yin, H. (2009). Exchange rate prediction using hybrid neural networks and trading indicators. Neurocomputing, 72(13-15), 2815-2823.
\bibitem{} Oh, K. J., \& Kim, K. J. (2002). Analyzing stock market tick data using piecewise nonlinear model. Expert Systems with Applications, 22(3), 249-255.
\bibitem{} Olson, D., \& Mossman, C. (2003). Neural network forecasts of Canadian stock returns using accounting ratios. International Journal of Forecasting, 19(3), 453-465.
\bibitem{Ou} Ou, J. A., \& Penman, S. H. (1989). Financial statement analysis and the prediction of stock returns. Journal of accounting and economics, 11(4), 295-329.
\bibitem{} Pai, P. F., \& Lin, C. S. (2005). A hybrid ARIMA and support vector machines model in stock price forecasting. Omega, 33(6), 497-505.
\bibitem{} Pan, H., Tilakaratne, C., \& Yearwood, J. (2005). Predicting Australian stock market index using neural networks exploiting dynamical swings and intermarket influences. Journal of research and practice in information technology, 37(1), 43.
\bibitem{} Panda, C., \& Narasimhan, V. (2007). Forecasting exchange rate better with artificial neural network. Journal of Policy Modeling, 29(2), 227-236.
\bibitem{} Pantazopoulos, K. N., Tsoukalas, L. H., Bourbakis, N. G., Brun, M. J., \& Houstis, E. N. (1998). Financial prediction and trading strategies using neurofuzzy approaches. IEEE Transactions on Systems, Man, and Cybernetics, Part B (Cybernetics), 28(4), 520-531.
\bibitem{} Petropoulos, A., Chatzis, S. P., Siakoulis, V., \& Vlachogiannakis, N. (2017). A stacked generalization system for automated FOREX portfolio trading. Expert Systems with Applications, 90, 290-302.
\bibitem{} Perez-Rodriguez, J. V., Torra, S., \& Andrada-Felix, J. (2005). STAR and ANN models: forecasting performance on the Spanish “Ibex-35” stock index. Journal of Empirical Finance, 12(3), 490-509.
\bibitem{} Qi, M. (1999). Nonlinear predictability of stock returns using financial and economic variables. Journal of Business \& Economic Statistics, 17(4), 419-429.
\bibitem{} Quah, T. S. (2008). DJIA stock selection assisted by neural network. Expert Systems with Applications, 35(1-2), 50-58.
\bibitem{} Quek, C., Yow, K. C., Cheng, P. Y., \& Tan, C. C. (2009). Investment portfolio balancing: application of a generic self‐organizing fuzzy neural network (GenSoFNN). Intelligent Systems in Accounting, Finance \& Management: International Journal, 16(1‐2), 147-164.
\bibitem{} Raposo, R. D. C. T., \& Cruz, A. D. O. (2002). Stock market prediction based on fundamentalist analysis with fuzzy-neural networks. In Proceedings of 3rd WSES International Conference on Fuzzy Sets.
\bibitem{} Rast, M. (1999, July). Forecasting with fuzzy neural networks: A case study in stock market crash situations. In Fuzzy Information Processing Society, 1999. NAFIPS. 18th International Conference of the North American (pp. 418-420). IEEE.
\bibitem{} Rather, A. M. (2011, December). A prediction based approach for stock returns using autoregressive neural networks. In Information and Communication Technologies (WICT), 2011 World Congress on (pp. 1271-1275). IEEE.
\bibitem{} Rather, A. M. (2014). A hybrid intelligent method of predicting stock returns. Advances in Artificial Neural Systems, 2014, 4.
\bibitem{} Rather, A. M., Agarwal, A., \& Sastry, V. N. (2015). Recurrent neural network and a hybrid model for prediction of stock returns. Expert Systems with Applications, 42(6), 3234-3241.
\bibitem{} Refenes, A. N., Azema-Barac, M., \& Zapranis, A. D. (1993). Stock ranking: Neural networks vs multiple linear regression. In Neural Networks, 1993., IEEE International Conference on (pp. 1419-1426). IEEE.
\bibitem{} Rodriguez-Gonzalez, A., Garcia-Crespo, A., Colomo-Palacios, R., Iglesias, F. G., \& Gomez-Berbis, J. M. (2011). CAST: Using neural networks to improve trading systems based on technical analysis by means of the RSI financial indicator. Expert Systems with Applications, 38(9), 11489-11500.
\bibitem{Rose} Rose, A. K., \& Stanley, T. D. (2005). A meta‐analysis of the effect of common currencies on international trade. Journal of economic surveys, 19(3), 347-365.
\bibitem{} Rout, A. K., Dash, P. K., Dash, R., \& Bisoi, R. (2017). Forecasting financial time series using a low complexity recurrent neural network and evolutionary learning approach. Journal of King Saud University-Computer and Information Sciences, 29(4), 536-552.
\bibitem{} Sagar, V. K., \& Kiat, L. C. (1999). A neural stock price predictor using qualitative and quantitative data. In Neural Information Processing, 1999. Proceedings. ICONIP'99. 6th International Conference on (Vol. 2, pp. 831-835). IEEE.
\bibitem{} Sezer, O. B., \& Ozbayoglu, A. M. (2018). Algorithmic Financial Trading with Deep Convolutional Neural Networks: Time Series to Image Conversion Approach. Applied Soft Computing.
\bibitem{} Shen, G., Tan, Q., Zhang, H., Zeng, P., \& Xu, J. (2018). Deep Learning with Gated Recurrent Unit Networks for Financial Sequence Predictions. Procedia computer science, 131, 895-903.
\bibitem{} Shynkevich, Y., McGinnity, T. M., Coleman, S. A., Belatreche, A., \& Li, Y. (2017). Forecasting price movements using technical indicators: Investigating the impact of varying input window length. Neurocomputing, 264, 71-88.
\bibitem{} Siekmann, S., Kruse, R., Gebhardt, J., Van Overbeek, F., \& Cooke, R. (2001). Information fusion in the context of stock index prediction. International journal of intelligent systems, 16(11), 1285-1298.
\bibitem{} Soto, J., \& Melin, P. (2015). Optimization of the interval type-2 fuzzy integrators in ensembles of ANFIS models for time series prediction: case of the Mexican stock exchange. In Design of Intelligent Systems Based on Fuzzy Logic, Neural Networks and Nature-Inspired Optimization (pp. 27-45). Springer, Cham.
\bibitem{Stanley} Stanley, T. D., \& Jarrell, S. B. (1989). Meta‐regression analysis: a quantitative method of literature surveys. Journal of economic surveys, 3, 161-170.
\bibitem{Stanley2} Stanley, T. , Doucouliagos, H. , Giles, M., Heckemeyer, J. H., Johnston, R. J., Laroche, P., Nelson, J. P., Paldam, M., Poot, J., Pugh, G., Rosenberger, R. S., \& Rost, K. (2013). Meta‐analysis of economics research reporting guidelines. Journal of Economic Surveys, 27(2), 390-394.
\bibitem{Steiner} Steiner, M., \& Wittkemper, H. G. (1997). Portfolio optimization with a neural network implementation of the coherent market hypothesis. European journal of operational research, 100(1), 27-40.
\bibitem{} Tay, F. E., \& Cao, L. (2001). Application of support vector machines in financial time series forecasting. Omega, 29(4), 309-317.
\bibitem{Tenti} Tenti, P. (1996). Forecasting foreign exchange rates using recurrent neural networks. Applied Artificial Intelligence, 10(6), 567-582.
\bibitem{} Thawornwong, S., \& Enke, D. (2004). The adaptive selection of financial and economic variables for use with artificial neural networks. Neurocomputing, 56, 205-232.
\bibitem{} Ticknor, J. L. (2013). A Bayesian regularized artificial neural network for stock market forecasting. Expert Systems with Applications, 40(14), 5501-5506.
\bibitem{Tsaih} Tsaih, R., Hsu, Y., \& Lai, C. C. (1998). Forecasting S\&P 500 stock index futures with a hybrid AI system. Decision Support Systems, 23(2), 161-174.
\bibitem{} Tseng, C. H., Cheng, S. T., Wang, Y. H., \& Peng, J. T. (2008). Artificial neural network model of the hybrid EGARCH volatility of the Taiwan stock index option prices. Physica A: Statistical Mechanics and its Applications, 387(13), 3192-3200.
\bibitem{} Vanstone, B. J., Finnie, G. R., \& Tan, C. N. (2005). Evaluating the Application of Neural Networks and Fundamental Analysis in the Australian Stockmarket. In Computational Intelligence (pp. 62-63).
\bibitem{} Versace, M., Bhatt, R., Hinds, O., \& Shiffer, M. (2004). Predicting the exchange traded fund DIA with a combination of genetic algorithms and neural networks. Expert Systems with applications, 27(3), 417-425.
\bibitem{Wah} Wah, B. W., \& Qian, M. (2002, July). Constrained formulations and algorithms for stock-price predictions using recurrent FIR neural networks. In AAAI/IAAI (pp. 211-216).
\bibitem{} Walczak, S. (1999). Gaining competitive advantage for trading in emerging capital markets with neural networks. Journal of Management Information Systems, 16(2), 177-192.
\bibitem{} Wang, J. L., \& Chan, S. H. (2006). Stock market trading rule discovery using two-layer bias decision tree. Expert Systems with Applications, 30(4), 605-611.
\bibitem{} Wang, J. Z., Wang, J. J., Zhang, Z. G., \& Guo, S. P. (2011). Forecasting stock indices with back propagation neural network. Expert Systems with Applications, 38(11), 14346-14355.
\bibitem{} Wang, Q., Xu, W., \& Zheng, H. (2018). Combining the wisdom of crowds and technical analysis for financial market prediction using deep random subspace ensembles. Neurocomputing, 299, 51-61.
\bibitem{} Wang, Y. H. (2009). Nonlinear neural network forecasting model for stock index option price: Hybrid GJR-GARCH approach. Expert Systems with Applications, 36(1), 564-570.
\bibitem{} Wen, Q., Yang, Z., \& Song, Y. (2009). Hybrid approaches for stock price prediction. Iccsa.
\bibitem{} Wen, Q., Yang, Z., Song, Y., \& Jia, P. (2010). Automatic stock decision support system based on box theory and SVM algorithm. Expert Systems with Applications, 37(2), 1015-1022.
\bibitem{} Witkowska, D. (1995). Neural networks as a forecasting instrument for the polish stock exchange. International Advances in Economic Research, 1(3), 232-241.
\bibitem{} Wittkemper, H. G., \& Steiner, M. (1996). Using neural networks to forecast the systematic risk of stocks. European Journal of Operational Research, 90(3), 577-588.
\bibitem{} Wu, X., Fung, M., \& Flitman, A. (2001, May). Forecasting stock market performance using hybrid intelligent system. In International Conference on Computational Science(pp. 447-456). Springer, Berlin, Heidelberg.
\bibitem{} Yeh, I. C., Lien, C. H., \& Tsai, Y. C. (2011). Evaluation approach to stock trading system using evolutionary computation. Expert Systems with Applications, 38(1), 794-803.
\bibitem{} Yu, L., Wang, S., \& Lai, K. K. (2008). Neural network-based mean-variance-skewness model for portfolio selection. Computers \& Operations Research, 35(1), 34-46.
\bibitem{} Yu, T. H. K., \& Huarng, K. H. (2008). A bivariate fuzzy time series model to forecast the TAIEX. Expert Systems with Applications, 34(4), 2945-2952.
\bibitem{} Yumlu, M. S., Gurgen, F. S., \& Okay, N. (2004). Turkish stock market analysis using mixture of experts. Proceedings of Engineering of Intelligent Systems (EIS), Madeira.
\bibitem{} Yumlu, S., Gurgen, F. S., \& Okay, N. (2005). A comparison of global, recurrent and smoothed-piecewise neural models for Istanbul stock exchange (ISE) prediction. Pattern Recognition Letters, 26(13), 2093-2103.
\bibitem{} Zhang, D., Jiang, Q., \& Li, X. (2004). Application of Neural Networks in Financial Data Mining. In International Conference on Computational Intelligence (pp. 392-395).
\bibitem{} Zhang, G. P. (2003). Time series forecasting using a hybrid ARIMA and neural network model. Neurocomputing, 50, 159-175.
\bibitem{} Zhang, G. P., \& Berardi, V. L. (2001). Time series forecasting with neural network ensembles: an application for exchange rate prediction. Journal of the Operational Research Society, 52(6), 652-664.
\bibitem{} Zhang, X. D., Li, A., \& Pan, R. (2016). Stock trend prediction based on a new status box method and AdaBoost probabilistic support vector machine. Applied Soft Computing, 49, 385-398.
\bibitem{} Zhang, Y. Q., \& Wan, X. (2007). Statistical fuzzy interval neural networks for currency exchange rate time series prediction. Applied Soft Computing, 7(4), 1149-1156.
\bibitem{} Zhang, Y. Q., Akkaladevi, S., Vachtsevanos, G., \& Lin, T. Y. (2002). Granular neural web agents for stock prediction. Soft Computing, 6(5), 406-413.
\bibitem{} Zhu, X., Wang, H., Xu, L., \& Li, H. (2008). Predicting stock index increments by neural networks: The role of trading volume under different horizons. Expert Systems with Applications, 34(4), 3043-3054.
\bibitem{} de Oliveira, F. A., Nobre, C. N., \& Zarate, L. E. (2013). Applying Artificial Neural Networks to prediction of stock price and improvement of the directional prediction index-Case study of PETR4, Petrobras, Brazil. Expert Systems with Applications, 40(18), 7596-7606.
\end{thebibliography}
\end{document}